\documentstyle[12pt,epsfig,psfig]{article}
\makeatletter

\@addtoreset{equation}{section}

\makeatother

\begin{document}

\title{The Casimir effect in the Fulling--Rindler vacuum}
\author{R. M. Avagyan, A. A. Saharian\footnote{%
E-mail address: saharyan@server.physdep.r.am}, A. H. Yeranyan\\
{\it Department of Physics, Yerevan State University,}\\ {\it 1
Alex Manoogian St., 375049 Yerevan, Armenia}}

\maketitle

\begin{abstract}
The vacuum expectation values of the energy--momentum tensor are
investigated for massless scalar fields satisfying Dicichlet or
Neumann boundary conditions, and for the electromagnetic field
with perfect conductor boundary conditions on two infinite
parallel plates moving by uniform proper acceleration through the
Fulling--Rindler vacuum. The scalar case is considered for general
values of the curvature coupling parameter and in an arbitrary
number of spacetime dimension. The mode--summation method is used
with combination of a variant of the generalized Abel--Plana
formula. This allows to extract manifestly the contributions to
the expectation values due to a single boundary. The vacuum forces
acting on the boundaries are presented as a sum of the
self--action and interaction terms. The first one contains well
known surface divergences and needs a further regularization. The
interaction forces between the plates are always attractive for
both scalar and electromagnetic cases. An application to the
'Rindler wall' is discussed.

\end{abstract}

\bigskip

PACS number(s): 03.70.+k, 11.10.Kk

\bigskip

\section{Introduction}

The imposition of boundary conditions on a quantum field leads to
the modification of the spectrum for the zero--point fluctuations
and results in the shift in the vacuum expectation values for
physical quantities such as the energy density and stresses. In
particular, vacuum forces arise acting on constraining boundaries.
This is the familiar Casimir effect. The particular features of
the resulting vacuum forces depend on the nature of the quantum
field, the type of spacetime manifold and its dimensionality, the
boundary geometries and the specific boundary conditions imposed
on the field. Since the original work by Casimir in 1948
\cite{Casimir} many theoretical and experimental works have been
done on this problem, including various types of boundary geometry
and non-zero temperature effects (see, e.g., \cite{Mostepanenko,
Plunien,Milton,Lamor,Bordag,Bordag1,Kirs01} and references
therein). Many different approaches have been used: mode summation
method with combination of the zeta function regularization
technique, Green function formalism, multiple scattering
expansions, heat-kernel series, etc. An interesting topic in the
investigations of the Casimir effect is the dependence of the
vacuum characteristics on the type of the vacuum. It is well known
that the uniqueness of vacuum state is lost when we work within
the framework of quantum field theory in a general curved
spacetime or in non--inertial frames. In particular, the use of
general coordinate transformation in quantum field theory in flat
spacetime leads to an infinite number of unitary inequivalent
representations of the commutation relations. Different
inequivalent representations will in general give rise to
different pictures with different physical implications, in
particular to different vacuum states. For instance, the vacuum
state for an uniformly accelerated observer, the Fulling--Rindler
vacuum \cite{Full73,Full77,Unru76,Boul75}, turns out to be
inequivalent to that for an inertial observer, the familiar
Minkowski vacuum. Quantum field theory in accelerated systems
contains many of special features produced by a gravitational
field avoiding some of the difficulties entailed by
renormalization in a curved spacetime. In particular, near the
canonical horizon in the gravitational field, a static spacetime
may be regarded as a Rindler--like spacetime. Note that, as it has
been shown in Ref. \cite{Avak01}, there is a class of solutions to
the Einstein equations with a plane--symmetric matter distribution
for which the corresponding external geometry is described by the
Rindler metric ('Rindler walls'). Another motivation for the
investigation of quantum effects in the Rindler space is related
to the fact that this space is conformally related to the de
Sitter space and to the Robertson--Walker space with negative
spatial curvature. As a result the expectation values of the
energy--momentum tensor for a conformally invariant field and for
corresponding conformally transformed boundaries on the de Sitter
and Robertson--Walker backgrounds can be derived from the
corresponding Rindler counterpart by the standard transformation
(see, for instance, \cite{Birrell}).

In this paper we will consider the vacuum expectation values of
the energy--momentum tensors for a scalar and electromagnetic
fields in the region between two parallel plates moving by
constant proper acceleration through the Fulling--Rindler vacuum.
This problem for a single plate case was considered by Candelas
and Deutsch \cite{Candelas} and by one of us \cite{Saharian1}. In
Ref. \cite{Candelas} the cases of conformally coupled Dirichlet
and Neumann massless scalar and electromagnetic fields are
investigated in the region of the right Rindler wedge on the right
from the barrier. In Ref. \cite{Saharian1} both regions, including
the one between the barrier and Rindler horizon are considered for
a massive scalar field with general curvature coupling parameter
and Robin boundary conditions in arbitrary number of spacetime
dimensions, and for the electromagnetic field. As in Ref.
\cite{Saharian1} (see also
\cite{Sahrev,RomSah,Rome01,Saha01,Reza02}), our regularization
scheme here is based on a variant of the generalized Abel--Plana
formula derived in Appendix \ref{section:App1}. This allows to
extract form the vacuum expectation values the single boundary
parts and to present the "interference" parts in terms of strongly
convergent integrals useful for numerical evaluations. We have
organized the paper as follows. In the next section we evaluate
the vacuum expectation values of the energy--momentum tensor for
the Dirichlet scalar. The corresponding interaction forces between
the plates are investigated in section \ref{sec:forces}. Section
\ref{sec:Neum} is dedicated to the case of the Neumann boundary
conditions. Then the vacuum densities and interaction forces for
the electromagnetic field are considered in section
\ref{sec:elmag}. Section \ref{sec:conc} concludes the main results
of the paper and an application to the 'Rindler wall' is
discussed. In Appendix \ref{section:App2} we consider the case of
the scalar field in two spacetime dimensions separately. An
alternate representation of the vacuum expectation values for the
energy--momentum tensor is obtained in Appendix
\ref{section:App3}.

\section{Vacuum energy-momentum tensor for a Dirichlet scalar}
\label{sec:Dir}

We consider a real massless scalar $\varphi (x)$ field with
general curvature coupling parameter $\zeta $ satisfying the field
equation
\begin{equation}
\nabla _{\mu }\nabla ^{\mu }\varphi +\zeta R\varphi =0,  \label{fieldeq}
\end{equation}
with $R$ being the scalar curvature for a $d+1$--dimensional
background spacetime, $\nabla _{\mu }$ is the covariant derivative
operator associated with the metric $g_{\mu \nu }$. For minimally
and conformally coupled scalars $\zeta =0$ and $\zeta =(d-1)/4d$,
respectively. By using field equation (\ref{fieldeq}) it can be
seen that the corresponding energy--momentum tensor (EMT) can be
presented in the form
\begin{equation}
T_{\mu \nu }=\nabla _{\mu }\varphi \nabla _{\nu }\varphi +\left[ \left(
\zeta -\frac{1}{4}\right) g_{\mu \nu }\nabla _{\rho }\nabla ^{\rho }-\zeta
\nabla _{\mu }\nabla _{\nu }-\zeta R_{\mu \nu }\right] \varphi ^{2},
\label{EMT}
\end{equation}
where $R_{\mu \nu }$ is the Ricci tensor.

Let $\{\varphi _\alpha (x),\varphi _\alpha ^{*}(x)\}$ is a
complete set of positive and negative frequency solutions to the
field equation (\ref{fieldeq}), where $\alpha $ denotes a set of
quantum numbers. Expanding field operator over these
eigenfunctions and using the commutation relations it can be
easily seen that the vacuum expectation values (VEV's) of the EMT
are presented in the form
\begin{equation}
\langle 0\mid T_{\mu \nu }\mid 0\rangle =\sum_\alpha T_{\mu \nu }\{\varphi
_\alpha ,\varphi _\alpha ^{*}\},  \label{EMTvev}
\end{equation}
where for a scalar field the quadratic form $T_{\mu \nu }\left\{ f
,g\right\} $ directly follows from the classical EMT given by Eq.
(\ref{EMT}).

\begin{figure}[tbph]
\begin{center}
\epsfig{figure=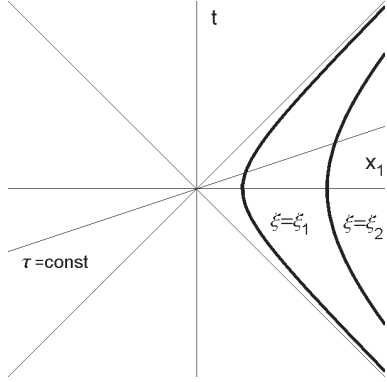,width=5cm,height=5cm}
\end{center}
\caption{The $(x^1,t)$ plane with the Rindler coordinates. The
heavy lines $\xi =\xi _1$ and $\xi =\xi _2$ represent the
trajectories of the plates.} \label{fig1avsa}
\end{figure}

Our main interest in this paper will be the vacuum expectation
values (VEV's) of the EMT in the Rindler spacetime induced by two
parallel plates moving with uniform proper acceleration when the
quantum field is prepared in the Fulling-Rindler vacuum. For this
problem the background spacetime is flat and in Eqs.
(\ref{fieldeq}),(\ref{EMT}) we have $R=0$, $R_{\mu \nu }=0$. As a
result the eigenmodes are independent on the curvature coupling
parameter and the EMT VEV's will depend on this parameter through
the expression (\ref{EMT}) only. In the following it will be
convenient to introduce Rindler coordinates $(\tau ,\xi ,{\bf x})$
related to the Minkowski ones, $(t,x^1,{\bf x})$ by
\begin{equation}
t=\xi \sinh \tau ,\quad x^1=\xi \cosh \tau ,  \label{RindMin}
\end{equation}
where ${\bf x}=(x^2,\ldots ,x^d)$ denotes the set of coordinates
parallel to the plates. In these coordinates the Minkowski line
element takes the form
\begin{equation}
ds^2=\xi ^2d\tau ^2-d\xi ^2-d{\bf x}^2,  \label{metric}
\end{equation}
and a wordline defined by $\xi ,{\mathbf{x}}={\mathrm{const}}$
describes an observer with constant proper acceleration $\xi
^{-1}$. Assuming that the plates are situated in the right Rindler
wedge $x^{1}>\left| t\right| $ we shall let the surfaces $\xi =\xi
_1$ and $\xi =\xi _2$, $\xi _2>\xi _1$ represent the trajectories
of these boundaries, which therefore have proper accelerations
$\xi _1^{-1}$ and $\xi _2^{-1}$ (see Fig. \ref{fig1avsa}). First
we will consider the case of a scalar field satisfying Dirichlet
boundary condition on the surface of the plates:
\begin{equation}
\varphi \mid _{\xi =\xi _{1}}=\varphi \mid _{\xi =\xi _{2}}=0
\label{Dboundcond}
\end{equation}
To evaluate the VEV's of the EMT by Eq. (\ref{EMTvev}) we need the
form of the eigenfunctions $\varphi _{\alpha }(x)$. For the
geometry under consideration the metric and boundary conditions
are static and translational invariant in the hyperplane parallel
to the plates. It follows from here that the corresponding part of
the eigenfunctions has the standard plane wave structure:
\begin{equation}
\varphi _{\alpha }=C\phi (\xi )\exp \left[ i\left( {\bf kx}-\omega
\tau \right) \right] ,\quad \alpha =({\bf k},\omega ),\quad
{\mathbf{k}}=(k_2,\ldots ,k_d). \label{wavesracture}
\end{equation}
The equation for $\phi (\xi )$ is obtained from field equation
(\ref {fieldeq}) on background of metric (\ref{metric}) and has
the form
\begin{equation}
\xi ^{2}\phi ^{\prime \prime }(\xi )+\xi \phi ^{\prime }(\xi )+\left( \omega
^{2}-k^{2}\xi ^{2}\right) \phi (\xi )=0,  \label{fiequ}
\end{equation}
where the prime denotes a differentiation with respect to the
argument, and $k=|{\bf k}|$. In the region between the plates the
corresponding linearly independent solutions to equation
(\ref{fiequ}) are the Bessel modified functions $I_{i\omega }(k\xi
)$ and $K_{i\omega }(k\xi )$. The solution satisfying boundary
condition (\ref{Dboundcond}) on the plate $\xi =\xi _{2}$ is in
form
\begin{equation}
D_{i\omega }(k\xi ,k\xi _{2})=I_{i\omega }(k\xi _{2})K_{i\omega }(k\xi
)-K_{i\omega }(k\xi _{2})I_{i\omega }(k\xi ).  \label{Deigfunc}
\end{equation}
Note that this function is real, $D_{i\omega }(k\xi ,k\xi
_{2})=D_{-i\omega }(k\xi ,k\xi _{2})$. From the boundary condition
on the plate $\xi =\xi _{1}$ we find that the possible values for
$\omega $ are roots to the equation
\begin{equation}
D_{i\omega }(k\xi _{1},k\xi _{2})=0.  \label{Deigfreq}
\end{equation}
This equation has an infinite set of solutions. We will denote
them by $\omega =\omega _{Dn}$, $\omega _{Dn}>0$, $n=1,2,\ldots $,
and will assume that they are arranged in the ascending order
$\omega _{Dn}<\omega _{Dn+1}$. The coefficient $C$ in formula
(\ref{wavesracture}) is determined from the standard Klein-Gordon
orthonormality condition for the eigenfunctions which for metric
(\ref{metric}) takes the form
\begin{equation}
\left( \varphi _{\alpha },\varphi _{\alpha ^{\prime }}\right) =
-i\int d{\mathbf x}\int _{\xi _1}^{\xi _2} \frac{d\xi }{\xi
}\varphi _{\alpha } \stackrel{\leftrightarrow }{\partial }_{\tau
}\varphi ^{*}_{\alpha '}= \delta _{\alpha \alpha ^{\prime }}.
\label{normcond}
\end{equation}
It can be easily seen that for any two solutions to equation
(\ref{fiequ}), $\phi _{\omega }^{(m)}(\xi )$, $m=1,2$ the
following integration formula takes place
\begin{equation}
\int_{\xi _{1}}^{\xi _{2}}\frac{d\xi }{\xi }\phi _{\omega
}^{(1)}(\xi ) \phi _{v}^{(2)}(\xi ) =\frac{\xi }{\omega
^{2}-\upsilon ^{2}}\left[ \phi _{\omega }^{(1)}(\xi )\frac{d\phi
_{\upsilon }^{(2)}(\xi )}{d\xi }-\phi _{\nu }^{(2)}(\xi
)\frac{d\phi _{\omega }^{(1)}(\xi )}{d\xi }\right] _{\xi
_{1}}^{\xi _{2}} . \label{intformula}
\end{equation}
Taking into account boundary condition (\ref{Dboundcond}) from Eq.
(\ref{normcond}) for the normalization coefficient one finds
\begin{equation}
C_{D}^{2}=\frac{1}{\left( 2\pi \right) ^{d-1}}\frac{I_{i\omega }(k\xi _{1})}{%
I_{i\omega }(k\xi _{2})\frac{\partial D_{i\omega }(k\xi _{1},k\xi _{2})}{%
\partial \omega }}\mid _{\omega =\omega _{Dn}}.  \label{Dnormc}
\end{equation}
Now substituting the eigenfunctions
\begin{equation}
\varphi _{\alpha }^{D}(x)=C_{D}D_{i\omega _{Dn}}(k\xi ,k\xi _{2})\exp \left[
i\left( {\bf kx}-\omega _{Dn}\tau \right) \right]  \label{Dsol2}
\end{equation}
into Eq. (\ref{EMTvev}) and integrating over the directions of $\
{\bf k}$ for the VEV's of the EMT we obtain diagonal form (no
summation over $i$)
\begin{equation}
\langle 0_D| T_{i}^{k}| 0_D\rangle =\delta _{i}^{k}\pi A_d
\int_{0}^{\infty }dkk^{d}\sum_{n=1}^{\infty }\frac{I_{i\omega
}(k\xi _{1})}{I_{i\omega }(k\xi
_{2})\frac{\partial D_{i\omega }(k\xi _{1},k\xi _{2})}{\partial \omega }}%
f^{(i)}[D_{i\omega }(k\xi ,k\xi _{2})]\mid _{\omega =\omega _{Dn}},
\label{EMTDdiag}
\end{equation}
where $|0_D\rangle $ is the amplitude for the Dirichlet vacuum
between the plates, and
\begin{equation}\label{Adnot}
  A_d=\frac{1}{2^{d-2}\pi ^{(d+1)/2}\Gamma (\frac{d-1}{2})}.
\end{equation}
In formula (\ref{EMTDdiag}) for a given function $G(z)$ we use the
notations
\begin{eqnarray}
f^{(0)}[G(z)] &=&\left( \frac{1}{2}-2\zeta \right) \left| \frac{dG(z)}{dz}%
\right| ^{2}+\frac{\zeta }{z}\frac{d}{dz}|G(z)|^{2}+\left[ \frac{1}{2}%
-2\zeta +\frac{\omega ^{2}}{z^{2}}\left( \frac{1}{2}+2\zeta \right) \right]
|G(z)|^{2},  \label{f0} \\
f^{(1)}[G(z)] &=&-\frac{1}{2}\left| \frac{dG(z)}{dz}\right| ^{2}-\frac{\zeta
}{z}\frac{d}{dz}|G(z)|^{2}+\frac{1}{2}\left( 1-\frac{\omega ^{2}}{z^{2}}%
\right) |G(z)|^{2},  \label{f1} \\
f^{(i)}[G(z)] &=&-\frac{|G(z)|^{2}}{d-1}-\left( 2\zeta -\frac{1}{2}\right) %
\left[ \left| \frac{dG(z)}{dz}\right| ^{2}+\left( 1-\frac{\omega ^{2}}{z^{2}}%
\right) |G(z)|^{2}\right] ;\quad i=2,\ldots ,d,  \label{f23}
\end{eqnarray}
where $G(z)=D_{i\omega }(z,k\xi _{2})$, and the indices 0,1
correspond to the coordinates $\tau $, $\xi $ respectively. It can
be easily seen that for a conformally coupled scalar the EMT
(\ref{EMTDdiag}) is traceless.

For the further evolution of VEV's (\ref{EMTDdiag}) we will apply
to the sum over $n$ summation formula (\ref{Dsumformula}) derived
in Appendix \ref{section:App1} by making use of the generalized
Abel-Plana formula \cite{Sahrev}. This yields
\begin{equation}
\langle 0_D| T_{i}^{k}|0_D\rangle = A_d \delta
_{i}^{k}\int_{0}^{\infty }dkk^{d}\int_{0}^{\infty }d\omega
\,\left\{ \frac{\sinh \pi \omega }{\pi }\,f^{(i)}[\tilde{D}
_{i\omega }(k\xi ,k\xi _{2})] - \frac{I_{\omega }(k\xi
_{1})}{I_{\omega }(k\xi _{2})}\frac{F^{(i)}[D_{\omega }(k\xi ,k\xi _{2})]}{%
D_{\omega }(k\xi _{1},k\xi _{2})}\right\} , \label{EMTDdiag1}
\end{equation}
where we have introduced the notation
\begin{equation}
\tilde{D}_{i\omega }(k\xi ,k\xi _{2})=K_{i\omega }(k\xi )-\frac{%
K_{i\omega }(k\xi _{2})}{I_{i\omega }(k\xi _{2})}I_{i\omega }(k\xi ),
\label{ztilda}
\end{equation}
and the functions $F^{(i)}[G(z)]$, $i=0,1,\ldots ,d$ are obtained
from the functions $f^{(i)}[G(z)]$ (see Eqs.
(\ref{f0})--(\ref{f23})) replacing $\omega \rightarrow i\omega $:
\begin{equation}
F^{(i)}[G(z)]=f^{(i)}[G(z),\omega \rightarrow i\omega ].  \label{Ffunc}
\end{equation}
The vacuum energy density, $\varepsilon $, effective pressures in
perpendicular, $p$, and parallel, $p_{\bot }$, to the plates
directions are determined by relations (no summation over $i$)
\begin{equation}
\varepsilon =\langle 0_D|T_{0}^{0}|0_D\rangle ,\quad p=-\langle
0_D| T_{1}^{1}|0_D\rangle ,\quad p_{\bot }=-\langle
0_D|T_{i}^{i}|0_D\rangle ,\quad i=2,\ldots ,d.  \label{enerpres}
\end{equation}
It can be easily checked from Eqs. (\ref{EMTDdiag1}),
(\ref{EMTNdiag1}) and (\ref {f0})--(\ref{f23}) that they satisfy
the standard continuity equation for the EMT, which for the
geometry under consideration takes the form
\begin{equation}
\frac{d(\xi p)}{d\xi }=-\varepsilon . \label{rel}
\end{equation}
For a conformally coupled scalar we have an additional zero--trace
relation $\varepsilon -p-(d-1)p_\perp =0$.
Let us consider the limit $\xi _{2}\rightarrow \infty $ of general formula (%
\ref{EMTDdiag1}) for fixed $\xi $. It can be easily seen that in
this limit the VEV's take the form
\begin{equation}
\lim_{\xi _{2}\rightarrow \infty }\langle 0_D| T_{i}^{k}|
0_D\rangle =\langle 0_{R}|T_{i}^{k}| 0_{R}\rangle +\langle
T_{i}^{k}\rangle _D^{(1b)}(\xi _{1},\xi ),\quad \xi
>\xi _{1}, \label{D1plater}
\end{equation}
where
\begin{equation}
\langle 0_{R}| T_{i}^{k}| 0_{R}\rangle =\frac{A_d\delta
_{i}^{k}}{\pi }\int_{0}^{\infty }dkk^{d}\int_{0}^{\infty }d\omega
\sinh \pi \omega \,f^{(i)}[K_{i\omega }(k\xi )]  \label{DFR}
\end{equation}
are the corresponding VEV's for the Fulling--Rindler vacuum
without boundaries, and the term
\begin{equation}
\langle T_{i}^{k}\rangle _D^{(1b)}(\xi _{1},\xi )=-A_d\delta
_{i}^{k}\int_{0}^{\infty }dkk^{d}\int_{0}^{\infty }d\omega
\frac{I_{\omega }(k\xi _{1})}{K_{\omega }(k\xi
_{1})}F^{(i)}[K_{\omega }(k\xi )]  \label{D1platebound}
\end{equation}
is induced in the region $\xi >\xi _1$ by the presence of a single
plane boundary located at $\xi =\xi _{1}$. Expressions
(\ref{D1platebound}) are finite for all values $\xi >\xi
_{1}$ and all divergences are contained in the purely Fulling-Rindler part (%
\ref{DFR}). These divergences can be regularized subtracting the
corresponding VEV's for the Minkowskian vacuum. The subtracted
VEV's
\begin{equation}\label{subRind}
  \langle T_{i}^{k}\rangle _{{\mathrm{sub}}}^{(R)}=\langle 0_R|T_{i}^{k}
  |0_R\rangle -\langle 0_M|T_{i}^{k}|0_M\rangle
\end{equation}
are investigated in a large number of papers (see, for instance,
\cite{Candelas,Saharian1,CandRaine,Davi77,Cand78,Troo79,Brow85,Brow86,%
Hill,Frol87,Dowk87,Pare93,More96} and references therein). The
most general case of a massive scalar field in an arbitrary number
of spacetime dimensions has been considered in Ref. \cite{Hill}
for conformally and minimally coupled cases and in Ref.
\cite{Saharian1} for general values of the curvature coupling
parameter (for the corresponding Green function see
\cite{CandRaine}). The formulae relevant to this paper are given
in \cite{Saharian1}. For a massless scalar VEV's (\ref{subRind})
can be presented in the form
\begin{equation}\label{subRindm0}
  \langle T_{i}^{k}\rangle _{{\mathrm{sub}}}^{(R)}=-\frac{\delta _i^k
  \xi ^{-d-1}}{2^{d-1}\pi ^{d/2}\Gamma (d/2)}\int _{0}^{\infty }
  \frac{\omega ^d g^{(i)}(\omega )d\omega }{e^{2\pi \omega }+(-1)^d}
\end{equation}
(the expressions for the functions $g^{(i)}(\omega )$ are given in
Ref. \cite{Saharian1}) correspond to the absence from the vacuum
of thermal distribution with standard temperature
$T=(2\pi\xi)^{-1}$. As we see from Eq. (\ref{subRindm0}), in
general, the corresponding spectrum has non-Planckian form: the
density of states factor is not proportional to $\omega
^{d-1}d\omega $. The spectrum takes the Planckian form for
conformally coupled scalars in $d=1,2,3$ with $g^{(0)}(\omega )=-d
g^{(i)}(\omega )=1$, $i=1,2,\ldots d$. It is interesting to note
that for even values of spatial dimension the distribution is
Fermi-Dirac type (see also \cite{Taga85,Oogu86}). For the massive
scalar the energy spectrum is not strictly thermal and the
corresponding quantities do not coincide with ones for the
Minkowski thermal bath.

The boundary induced quantities (\ref{D1platebound}) are
investigated in Ref. \cite{Candelas} for a conformally coupled
$d=3$ massless Dirichlet scalar and in Ref. \cite{Saharian1} for a
massive scalar with general curvature coupling and Robin boundary
condition in an arbitrary number of dimensions. The single
boundary part (\ref{D1platebound}) diverges at the plate surface
$\xi =\xi _1$ with leading terms proportional to $(\xi -\xi
_1)^{-d-1}$ for $i=0,2,\ldots ,d$ and to $(\xi -\xi _1)^{-d}$ for
$i=1$ (see below). These leading terms vanish for a conformally
coupled scalar, and for $i=0,2,\ldots ,d$ coincide with the
corresponding quantities for a plane boundary in the Minkowski
vacuum \cite{Saharian1}.

Now we turn to the limit $\xi _{1}\rightarrow 0$ in formula
(\ref{EMTDdiag1}), when the left plate coincides with the right
Rindler horizon. In this limit in the second
term on the right of formula (\ref{EMTDdiag1}) the subintegrand behaves as $%
\xi _{1}^{2\omega }$ and tends to zero. As a result one obtains
\begin{equation}
\lim_{\xi _{1}\rightarrow 0}\langle 0_D| T_{i}^{k}| 0_D\rangle =%
\frac{A_d\delta _{i}^{k}}{\pi } \int_{0}^{\infty
}dkk^{d}\int_{0}^{\infty }d\omega \,\sinh \pi \omega
\,f^{(i)}[\tilde{D}_{i\omega }(k\xi ,k\xi _{2})]. \label{singleft}
\end{equation}
These quantities coincide with the corresponding ones induced in
the region $\xi <\xi _{2}$ by a single plate at $\xi =\xi _{2}$.
They are investigated in Ref. \cite{Saharian1}, where it has been
shown that the VEV's (\ref{singleft}) can be presented in the form
similar to Eq. (\ref{D1plater}):
\begin{equation}
\lim_{\xi _{1}\rightarrow 0}\langle 0_D| T_{i}^{k}| 0_D\rangle
=\langle 0_{R}| T_{i}^{k}| 0_{R}\rangle +\langle T_{i}^{k}\rangle
^{(1b)}_D(\xi _{2},\xi ),\quad \xi <\xi _{2}, \label{D1platel}
\end{equation}
where the expressions for the boundary part $\langle T_{i}^{
k}\rangle _D^{(1b)}(\xi _{2},\xi )$ in the region $\xi <\xi _{2}$
are obtained from formulae (\ref{D1platebound}) by replacing (see
Ref. \cite{Saharian1})
\begin{equation}
I_{\omega }\rightarrow K_{\omega },\quad K_{\omega }\rightarrow I_{\omega
},\quad \xi _{1}\rightarrow \xi _{2},\quad \xi _{2}\rightarrow \xi _{1}.
\label{replace}
\end{equation}
By using Eqs. (\ref{EMTDdiag1}),(\ref{singleft}),(\ref{D1platel}) the parts
in the VEV's induced by the existence of boundaries,
\begin{equation}
\langle T_{i}^{k}\rangle ^{(b)}_D=\langle 0_D| T_{i}^{k}|
0_D\rangle -\langle 0_{R}| T_{i}^{k}| 0_{R}\rangle ,
\label{Tikbound}
\end{equation}
can be written as
\begin{equation}
\langle T_{i}^{}\rangle ^{(b)}_D\left( \xi _{1},\xi _{2},\xi
\right) =\langle T_{i}^{k}\rangle _D^{(1b)}\left( \xi _{2},\xi
\right) -A_d \delta _{i}^{k} \int_{0}^{\infty }dk\,k^{d}
\int_{0}^{\infty }d\omega \frac{I_{\omega }(k\xi _{1})}{I_{\omega
}(k\xi _{2})}\frac{F^{(i)}[D_{\omega }(k\xi ,k\xi
_{2})]}{D_{\omega }(k\xi _{1},k\xi _{2})}. \label{Tikbound1}
\end{equation}

In Appendix C we show that the VEV's (\ref{EMTDdiag1}) can be also
presented in the form (\ref{EMT1diag1}). Substituting Eq.
(\ref{limcond}) into this formula, the boundary VEV's can be also
written in the form
\begin{equation}
\langle T_{i}^{k}\rangle ^{(b)}_D\left( \xi _{1},\xi _{2},\xi
\right) =\langle T_{i}^{k}\rangle ^{(1b)}_D(\xi _{1},\xi
)-A_d\delta _{i}^{k}\int_{0}^{\infty }dk\,k^{d} \int_{0}^{\infty
}d\omega \frac{K_{\omega }(k\xi _{2})}{K_{\omega }(k\xi
_{1})}\frac{F^{(i)}[D_{\omega }(k\xi ,k\xi _{1})]}{D_{\omega
}(k\xi _{1},k\xi _{2})}. \label{EMTDform2}
\end{equation}
This expression is obtained from Eq. (\ref{Tikbound1}) by
replacements (\ref{replace}). The case $d=1$ needs a separate
consideration and is investigated in Appendix \ref{section:App2}.
It can be seen that the corresponding formulae for the VEV's are
also obtained from the formulae given above in this section
replacing
\begin{equation}\label{d1replace}
A_d\int _{0}^{\infty }dk\, k^{d-2}\rightarrow \frac{1}{\pi },\quad
k\rightarrow 0.
\end{equation}

Now let us present the VEV's (\ref{EMTDdiag1}) in the form
\begin{equation}
\langle 0| T_{i}^{k}| 0\rangle _D=\langle 0_{R}| T_{i}^{k}|
0_{R}\rangle +\langle T_{i}^{k}\rangle ^{(1b)}_D(\xi _{1},\xi
)+\langle T_{i}^{k}\rangle ^{(1b)}_D(\xi _{2},\xi )+\Delta \langle
T_{i}^{k}\rangle _D(\xi _{1},\xi _{2},\xi ),\quad \xi _{1}<\xi
<\xi _{2},
\end{equation}
where
\begin{equation} \label{intterm1}
\Delta \langle T_{i}^{k}\rangle _D=-A_d\delta _i^k\int_0^\infty dk
k^d\int_0^\infty d\omega
I_\omega (k\xi _1)\left[ \frac{F^{(i)}[D_\omega (k\xi ,k\xi _2)]}{%
I_\omega (k\xi _2)D_\omega (k\xi _1,k\xi
_2)}-\frac{F^{(i)}[K_\omega (k\xi )]}{K_\omega (k\xi _1)}\right]
\end{equation}
is the 'interference' term. The surface divergences are contained
in the single boundary parts and this term is finite for all
values $\xi _1\leq \xi \leq \xi _2$. An equivalent formula for
$\Delta \langle T_{i}^{k}\rangle _D$ is obtained from Eq.
(\ref{intterm1}) by replacements (\ref{replace}). In the limit
$\xi _1\to \xi _2$ expressions (\ref{intterm1}) are divergent and
for small values of $\xi_2/\xi _1-1$ the main contribution comes
from the large values of $\omega $. Introducing a new integration
variable $x=k/\omega $ and replacing Bessel modified functions by
their uniform asymptotic expansions for large values of the order
(see Ref. \cite{Abramowitz}) at the leading order over $1/(\xi
_2-\xi _1)$ one receives (no summation over $i$)
\begin{eqnarray}\label{1basymp}
  \langle T_i^i\rangle ^{(1b)}_D(\xi _j,\xi )&\sim & \frac{d(\zeta _c-
  \zeta )\Gamma \left( \frac{d+1}{2}\right)}{2^d\pi ^{(d+1)/2}|\xi -\xi _j|
  ^{d+1}},\quad i=0,2,\ldots ,d ,\\
  \langle T_1^1\rangle ^{(1b)}_D(\xi _j,\xi )&\sim & \langle T_0^0
  \rangle ^{(1b)}_D(\xi _j,\xi )\frac{\xi _j-\xi }{d\xi _j},\quad
  j=1,2
  \label{1basymppe}
\end{eqnarray}
for the single boundary terms, and
\begin{eqnarray}\label{D2Mink0}
  \Delta \langle T_{0}^{0}\rangle _D& \sim & -\frac{1}{d}
  \Delta \langle T_{1}^{1}\rangle _D +\frac{(\zeta -\zeta _c)
  (\xi _2-\xi _1)^{-d-1}}{2^{2d-1}\pi ^{d/2}\Gamma (d/2)} \times
  \\
  && \times \int _{0}^{\infty }\frac{dt t^{d}}{e^t-1}\left[ \exp \left( t
  \frac{\xi _1-\xi }{\xi _2-\xi _1}\right)+
  \exp \left( t\frac{\xi  -\xi _2 }{\xi _2-\xi _1}\right)\right]
  \nonumber \\
\Delta \langle T_{1}^{1}\rangle _D&\sim & \frac{d\zeta _R(d+1)
\Gamma \left(\frac{d+1}{2}\right)}{(4\pi )^{(d+1)/2}(\xi _2-\xi
_1)^{d+1}},\quad \Delta \langle T_{i}^{i}\rangle _D\sim \Delta
\langle T_{0}^{0}\rangle _D, \quad i=2,3,\ldots , \label{D2Mink1}
\end{eqnarray}
for the 'interference' terms. Here $\zeta _R(s)$ is the Riemann
zeta--function. Expressions (\ref{1basymp}), (\ref{D2Mink0}),
(\ref{D2Mink1}) coincide with the corresponding formulae for two
parallel plates geometry in $d+1$ -- dimensional Minkowski
spacetime with separation $\xi _2-\xi _1$ (see Ref. \cite{Amb83}
for the conformally coupled case and Ref. \cite{RomSah} for the
general case of the curvature coupling parameter $\zeta $). Note
that in the limit under consideration the 'interference' term
(\ref{D2Mink1}) for the vacuum perpendicular pressure dominates
the single boundary induced terms, given by Eq. (\ref{1basymppe}).

\section{Interaction forces between the plates} \label{sec:forces}

Now we turn to the interaction forces between the plates. The
vacuum force acting per unit surface of the plate at $\xi =\xi
_{i}$ is determined by the ${}^{1}_{1}$--component of the vacuum
EMT at this point. The corresponding effective pressures can be
presented as a sum of two terms:
\begin{equation}
p_{D}^{(i)}=p_{D1}^{(i)}+p_{D{\rm (int)}}^{(i)},\quad i=1,2.
\label{FintD}
\end{equation}
The first term on the right is the pressure for a single plate at $%
\xi =\xi _{i}$ when the second plate is absent. This term is
divergent due to the well known surface divergences in the
subtracted VEV's. The second term on the right of Eq.
(\ref{FintD}),
\begin{equation}
p_{D{\rm (int)}}^{(i)}=-\langle T_{1}^{1}\rangle ^{(1b)}_D(\xi
_{j},\xi _{i})-\Delta \langle T_{1}^{1}\rangle _D(\xi _{1},\xi
_{2},\xi _{i}),\quad i,j=1,2,\quad j\neq i \label{pintD}
\end{equation}
is the pressure induced by the presence of the second plate, and
can be termed as an interaction force. For the plate at $\xi =\xi
_{2}$ the interaction term is due to the second summand on the
right of Eq. (\ref{EMTDdiag1}). Substituting into this term $\xi
=\xi _{2}$ and using the Wronskian relation for the modified
Bessel functions one has
\begin{equation}
p_{D{\rm (int)}}^{(2)}(\xi _1,\xi _2)=-\frac{A_d}{2\xi _{2}^{2}}
\int_{0}^{\infty }dkk^{d-2}\int_{0}^{\infty }d\omega \frac{%
I_{\omega }(k\xi _{1})}{I_{\omega }(k\xi _{2})D_{\omega }(k\xi _{1},k\xi
_{2})}.  \label{pint2}
\end{equation}
By a similar way from Eq. (\ref{EMTDform2}) for the interaction
term on the plate at $\xi =\xi _{1}$ we obtain
\begin{equation}
p_{D{\rm (int)}}^{(1)}(\xi _1,\xi _2)=-\frac{A_d}{2\xi _{1}^{2}}
\int_{0}^{\infty }dkk^{d-2}\int_{0}^{\infty }d\omega \frac{%
K_{\omega }(k\xi _{2})}{K_{\omega }(k\xi _{1})D_{\omega }(k\xi _{1},k\xi
_{2})}.  \label{pint1}
\end{equation}
As the function $D_{\omega }(k\xi ,k\xi _2)$ is positive for $\xi
_1<\xi _2$, interaction forces per unit surface (\ref{pint2}) and
(\ref{pint1}) are always attractive. They are finite for all $\xi
_{1}<\xi _{2}$, and do not depend on the curvature coupling
parameter $\zeta $. In the limit $\xi _{1}\rightarrow \xi
_{2}$ these forces diverge due the contribution from the large values $%
\omega $ and in this limit by introducing a new integration
variable we can replace the Bessel modified functions by their
uniform asymptotic expansions for large values of the order. At
the leading order for the perpendicular vacuum pressures we obtain
formula (\ref{D2Mink1}) which corresponds to the standard Casimir
attraction force for two parallel plates in Minkowski vacuum.

From expressions (\ref{pint2}) and (\ref{pint1}) it follows that
\begin{equation}\label{pint12ineq}
p_{D{\rm (int)}}^{(2)}(\xi _1,\xi _2)>p_{D{\rm (int)}}^{(1)}(\xi
_1,\xi _2).
\end{equation}
This can be proved by using that the function $z^2I_{\omega
}(z)K_{\omega }(z)$ is monotonic increasing. The latter directly
follows from the relations
\begin{eqnarray}\label{Iineq1}
  && \sqrt{1+\frac{(\omega +1)^2}{z^2}}-\frac{1}{z}<\frac{I'_{\omega }
  (z)}{I_{\omega}(z)}<\sqrt{1+\frac{\omega ^2}{z^2}} \\
  && \sqrt{1+\frac{(\omega +1)^2}{z^2}}+\frac{1}{z}>-\frac{K'_{\omega }
  (z)}{K_{\omega}(z)}>\sqrt{1+\frac{\omega ^2}{z^2}}. \label{Kineq2}
\end{eqnarray}
The proof for the right inequalities in Eqs.
(\ref{Iineq1}),(\ref{Kineq2}) is presented in Ref.
\cite{Candelas}. The left inequalities are obtained from the
recurrence relations for the Bessel modified functions. For
instance, in the case of the function $I_{\omega }(z)$ one has:
\begin{eqnarray}\label{proofineq1}
  && \frac{I'_{\omega }(z)}{I_{\omega}(z)}= \frac{I_{\omega +1}
  (z)}{I_{\omega}(z)}+\frac{\omega }{z}=\left[ \frac{I'_{\omega +1}
  (z)}{I_{\omega +1}(z)}+\frac{\omega +1}{z}\right] ^{-1}+\frac{\omega
  }{z}> \nonumber \\
  && >\left[ \sqrt{1+\frac{(\omega +1)^2}{z^2}}+\frac{\omega +
  1}{z}\right] ^{-1}+\frac{\omega }{z}=\sqrt{1+\frac{(\omega
  +1)^2}{z^2}}-\frac{1}{z},
\end{eqnarray}
where we have used the right inequality in Eq. (\ref{Iineq1}). The
left inequality in Eq. (\ref{Kineq2}) can be proved in a similar
way.

To see the monotonicity properties of functions (\ref{pint2}) and
(\ref{pint1}) note that
\begin{equation}
\xi _{1}\frac{\partial p_{D{\rm (int)}}^{(1)}}{\partial \xi _{2}}=-\xi _{2}%
\frac{\partial p_{D{\rm (int)}}^{(2)}}{\partial \xi
_{1}}=\frac{A_{d}}{2\xi
_{1}\xi _{2}}\int_{0}^{\infty }dkk^{D-2}\int_{0}^{\infty }\frac{d\omega }{%
D_{\omega }^{2}(k\xi _{1},k\xi _{2})}.  \label{pDxi2}
\end{equation}
It follows from here that for a fixed value of $\xi _{1}$ ($\xi
_{2}$) the quantity $p_{D{\rm (int)}}^{(1)}$ ($p_{D{\rm
(int)}}^{(2)}$) is monotonic increasing (decreasing) function on
$\xi _{2}$ ($\xi _{1}$). By taking into account that both this
quantities are negative we conclude that the modulus of the
interaction force on the plate at $\xi _{1}$ ($\xi _{2}$) is
monotonic decreasing (increasing) function on $\xi _{2}$ ($\xi
_{1}$) for a fixed value of $\xi _{1}$ ($\xi _{2}$). From formula
(\ref{pint2}) it follows that
\begin{equation}
\xi _{i}\frac{\partial p_{D{\rm (int)}}^{(i)}}{\partial \xi _{i}}=-(d+1)p_{D%
{\rm (int)}}^{(i)}-\xi _{j}\frac{\partial p_{D{\rm
(int)}}^{(i)}}{\partial \xi _{j}},\quad i,j=1,2,\quad i\neq j.
\label{pDxi22}
\end{equation}
For $i=2$ the both terms on the right are positive and hence, the
same is the case for the function on the left. Therefore for a
fixed $\xi _{1}$ the function $p_{D{\rm (int)}}^{(2)}$ is
monotonic increasing on $\xi _{2}$ and the modulus of the
corresponding interaction force is monotonic decreasing function
on $\xi _{2}$. In the case $i=1$ the terms on the right in this
formula have different signs. For a fixed value of $\xi _{2}$ the function $%
p_{D{\rm (int)}}^{(1)}$ is monotonic increasing on $\xi _{1}$ near
the horizon, $\xi _{1}\rightarrow 0$, and monotonic decreasing
near the second plane, $\xi _{1}\rightarrow \xi _{2}$. It follows
from here the modulus of the corresponding interaction force
acting on the plate at $\xi _{1}$ has minimum for some
intermediate value.

\begin{figure}[tbph]
\begin{center}
\epsfig{figure=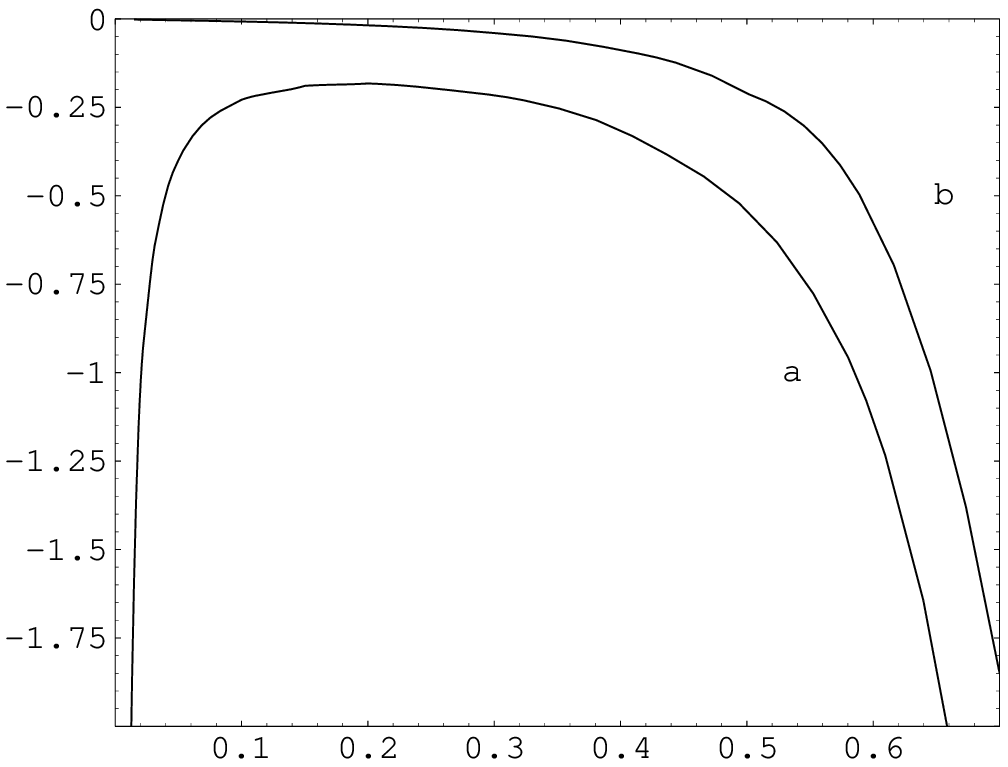,width=7cm,height=5cm}
\end{center}
\caption{ The $d=3$ vacuum effective pressures determining the
interaction forces between Dirichlet parallel plates, multiplied
by $\xi ^{4}_2$, $\xi ^{4}_2 p_{D{\rm (int)}}^{(1)}$ (curve a) and
$\xi ^{4}_2 p_{D{\rm (int)}}^{(2)}$ (curve b) as functions of the
ratio $\xi _1/\xi _2 $.} \label{fig2pintD}
\end{figure}

In the limit $\xi _2\gg \xi _1$, introducing in Eq. (\ref{pint2})
a new integration variable $x=k\xi _2$, and making use the formula
\begin{equation}\label{Ismall}
  I_\omega (y)=\left( \frac{y}{2}\right) ^{\omega }\frac{1}{\Gamma (\omega )}
  \left[ 1+{\cal O}(y^2)\right] ,\quad y=x\xi _1/\xi _2,
\end{equation}
and the standard relation between the functions $K_\omega $ and
$I_{\pm \omega }$ one finds
\begin{equation}
p_{D{\rm (int)}}^{(2)} \approx -\frac{\pi ^2A_d}{48 \xi
_{2}^{d+1}\ln ^2(2\xi _{2}/\xi _{1})}\int_{0}^{\infty
}\frac{dxx^{d-2}}{I_{0}^{2}(x)}\left[1+{\cal O}\left( \frac{\ln x
}{\ln (2\xi _2/\xi _1)} \right) \right] . \label{pD2far}
\end{equation}
The similar calculation for Eq. (\ref{pint1}) yields
\begin{equation}
p_{D{\rm (int)}}^{(1)} \approx -\frac{\pi ^2A_d}{24 \xi
_{2}^{d-1}\xi ^2_1\ln ^3(2\xi _{2}/\xi _{1})}\int_{0}^{\infty
}\frac{dxx^{d-2}K_0(x)}{I_{0}(x)}\left[1+{\cal O}\left( \frac{\ln
x }{\ln (2\xi _2/\xi _1)} \right) \right] . \label{pD1far}
\end{equation}
We have carried out numerical evaluations for the interaction
forces by making use of formulae (\ref{pint2}) and (\ref{pint1}).
In Fig. \ref{fig2pintD} the corresponding results are presented
for $\xi _2^{d+1}p_{D{\rm (int)}}^{(i)}$, $i=1,2$ in the case
$d=3$ as functions on $\xi _1/\xi _2$.

\section{VEV's and the interaction forces for the Neumann scalar}
\label{sec:Neum}

In this section we will consider VEV's for the EMT in the case of a scalar
field satisfying the Neumann boundary condition on the plates $\xi =\xi
_{1},\xi _{2}$:
\begin{equation}
\frac{\partial \varphi }{\partial \xi }| _{\xi =\xi
_{1}}=\frac{\partial \varphi }{\partial \xi }| _{\xi =\xi _{2}}=0.
\label{Nboundcond}
\end{equation}
The corresponding scheme is similar to that given above for the
Dirichlet case. The eigenfunctions to the field equation
(\ref{fieldeq}) have form (\ref{wavesracture}) with
\begin{equation}
\phi (\xi )=N_{i\omega }(k\xi ,k\xi _{2})=I_{i\omega }^{\prime }(k\xi
_{2})K_{i\omega }(k\xi )-K_{i\omega }^{\prime }(k\xi _{2})I_{i\omega }(k\xi
).  \label{Neigfunc}
\end{equation}
As in the Dirichlet case this function is real. From the boundary condition
on the plate $\xi =\xi _{1}$ we obtain that the corresponding
eigenfrequencies are solutions to the equation
\begin{equation}
N'_{i\omega }(k\xi _1,k\xi _{2})=I_{i\omega }^{\prime }(k\xi _{2})
K_{i\omega }^{\prime }(k\xi _{1})-K_{i\omega }^{\prime }(k\xi
_{2}) I_{i\omega }^{\prime }(k\xi _{1})=0. \label{Neigfreq}
\end{equation}
We will denote them by ${\omega =\omega _{Nn}}$, $n=1,2,...$ ,
arranged in the ascending order $\omega _{Nn}<\omega _{Nn+1}$. The
normalization coefficient $C$ can be found from orthonormality
condition (\ref{normcond}) using integration formula
(\ref{intformula}):
\begin{equation}
C_N^2=\frac 1{\left( 2\pi \right) ^{d-1}}\frac{I_{i\omega
}^{\prime }(k\xi _1)}{I_{i\omega }^{\prime }(k\xi
_2)\frac{\partial N_{i\omega }^{\prime }(k\xi _1,k\xi
_2)}{\partial \omega }}| _{\omega =\omega _{Nn}}. \label{Nnormc}
\end{equation}
Substituting the eigenfunctions into the mode sum formula (\ref{EMTvev}) one
obtains
\begin{equation}
\langle 0_N|T_{i}^{k}| 0_N\rangle =\pi A_d\delta _i^k
\int_0^\infty dkk^d\sum_{n=1}^\infty \frac{I_{i\omega }^{\prime
}(k\xi _1)}{I_{i\omega
}^{\prime }(k\xi _2)\frac{\partial N_{i\omega }^{\prime }(k\xi _1,k\xi _2)}{%
\partial \omega }}f^{(i)}[N_{i\omega }(k\xi ,k\xi _2)]| _{\omega =\omega
_{Nn}},  \label{EMTNdiag}
\end{equation}
where $|0_N\rangle $ is the amplitude for the Neumann vacuum state
between the plates, and the functions $f^{(i)}[G(z)]$ are defined
in accordance with Eqs. (\ref{f0})--(\ref{f23}). To sum the series
over the eigenfrequencies $ \omega _{Nn}$ we will apply the
summation formula derived in Appendix A, Eq. (\ref{Nsumformula}).
This yields
\begin{equation}
\langle 0_N| T_{i}^{k}| 0_N\rangle =A_d\delta _i^k\int_0^\infty
dkk^d\int_0^\infty d\omega \,\left\{\frac{\sinh \pi \omega }{\pi }
\,f^{(i)}[\tilde{N}_{i\omega }(k\xi ,k\xi _2)]- \frac{I_\omega
^{\prime }(k\xi _1)}{I_\omega ^{\prime }(k\xi
_2)}\frac{F^{(i)}[N_\omega (k\xi ,k\xi _2)]}{N'_{i\omega }(k\xi
_1,k\xi _{2})}\right\} , \label{EMTNdiag1}
\end{equation}
with functions $F^{(i)}[G(z)]$ defined as in Eq. (\ref{Ffunc}),
and we use the notation
\begin{equation}
\tilde{N}_\omega (k\xi ,k\xi _2)=K_{i\omega }(k\xi
)-\frac{K_{i\omega }^{\prime }(k\xi _2)}{I_{i\omega }^{\prime
}(k\xi _2)}I_{i\omega }(k\xi ). \label{Ntilde}
\end{equation}
To identify the terms in Eq. (\ref{EMTNdiag1}) let us consider
limiting cases. In the limit $\xi _2\to \infty $, from Eq.
(\ref{EMTNdiag1}) one obtains
\begin{equation}
\lim_{\xi _{2}\to \infty }\langle 0_N| T_{i}^{k}| 0_N\rangle
=\langle 0_{R}|T_{i}^{k}| 0_{R}\rangle +\langle T_{i}^{k}\rangle
^{(1b)}_N(\xi _{1},\xi ), \label{N1plater}
\end{equation}
where the term
\begin{equation}
\langle T_{i}^{k}\rangle ^{(1b)}_N(\xi _{1},\xi )=-A_d\delta
_{i}^{k}\int_{0}^{\infty }dkk^{d}\int_{0}^{\infty }d\omega
\frac{I'_{\omega }(k\xi _{1})}{K'_{\omega }(k\xi
_{1})}F^{(i)}[K_{\omega }(k\xi )]  \label{N1platebound}
\end{equation}
is induced in the region $\xi >\xi _1 $ by a single Neumann
boundary located at $\xi =\xi _{1}$. This quantities for $d=3$
case are investigated in Ref. \cite{Candelas}. In the limit $\xi
_1\to 0$ the left plate coincides with the Rindler horizon and the
second term in the figure braces in Eq. (\ref{EMTNdiag1})
vanishes. In this case the VEV's coincide with the corresponding
expressions for a single plate at $\xi =\xi _{2}$ induced in the
region $\xi <\xi _{2}$. They are investigated in Ref.
\cite{Saharian1}, where it has been shown that the VEV's
(\ref{singleft}) can be presented in the form similar to Eq.
(\ref{N1plater}):
\begin{equation}
\lim_{\xi _{1}\to 0}\langle 0_N| T_{i}^{k}| 0_N\rangle =\langle
0_{R}| T_{i}^{k}| 0_{R}\rangle +\langle T_{i}^{k}\rangle
^{(1b)}_N(\xi _{2},\xi ),\quad \xi <\xi _{2}, \label{N1platel}
\end{equation}
where the expressions for the boundary part $\langle T_{Ni}^{\quad
k}\rangle ^{(1b)}_N(\xi _{2},\xi )$ in the region $\xi <\xi _{2}$
are obtained from formulae (\ref{N1platebound}) by replacements
(\ref{replace}).

By using Eqs. (\ref{EMTNdiag1}),(\ref{N1platel}) the parts in the
VEV's induced by the existence of boundaries,
\begin{equation}
\langle T_{i}^{k}\rangle ^{(b)}_N=\langle 0_N| T_{i}^{k}|
0_N\rangle -\langle 0_{R}| T_{i}^{k}| 0_{R}\rangle ,
\label{TikboundN}
\end{equation}
can be presented as
\begin{equation}
\langle T_{i}^{k}\rangle ^{(b)}_N\left( \xi _{1},\xi _{2},\xi
\right) =\langle T_{i}^{k}\rangle ^{(1b)}_N\left( \xi _{2},\xi
\right) -A_d \delta _{i}^{k} \int_{0}^{\infty }dk\,k^{d}
\int_{0}^{\infty }d\omega \frac{I'_{\omega }(k\xi
_{1})}{I'_{\omega }(k\xi _{2})}\frac{F^{(i)}[N_{\omega }(k\xi
,k\xi _{2})]}{N'_{\omega }(k\xi _{1},k\xi _{2})}.
\label{Tikbound1N}
\end{equation}
Similar to the Dirichlet case, the Neumann boundary VEV's can be
also written in the form
\begin{equation}
\langle T_{i}^{k}\rangle ^{(b)}_N\left( \xi _{1},\xi _{2},\xi
\right) =\langle T_{i}^{k}\rangle ^{(1b)}_N(\xi _{1},\xi
)-A_d\delta _{i}^{k}\int_{0}^{\infty }dk\,k^{d}\int_{0}^{\infty
}d\omega \frac{K'_{\omega }(k\xi _{2})}{K'_{\omega }(k\xi
_{1})}\frac{F^{(i)}[N_{\omega }(k\xi ,k\xi _{1})]}{N'_{\omega
}(k\xi _{1},k\xi _{2})}, \label{EMTNform2}
\end{equation}
with $\langle T_{i}^{k}\rangle ^{(1b)}_N(\xi _{1},\xi )$ being the
VEV's induced by a single Neumann boundary located at $\xi =\xi
_1$. As we see, this expression is obtained from
(\ref{Tikbound1N}) by replacements (\ref{replace}).

Now let us present the VEV's (\ref{EMTNdiag1}) in the form
\begin{equation}
\langle 0_N| T_{i}^{k}| 0_N\rangle =\langle 0_{R}| T_{i}^{k}|
0_{R}\rangle +\langle T_{i}^{k}\rangle ^{(1b)}_N(\xi _{1},\xi
)+\langle T_{i}^{k}\rangle ^{(1b)}_N(\xi _{2},\xi )+\Delta \langle
T_{i}^{k}\rangle _N(\xi _{1},\xi _{2},\xi ),\quad \xi _{1}<\xi
<\xi _{2},
\end{equation}
where
\begin{equation} \label{intterm1N}
\Delta \langle T_{i}^{k}\rangle _N=-A_d\delta _i^k\int_0^\infty dk
k^d\int_0^\infty d\omega
I'_\omega (k\xi _1)\left[ \frac{F^{(i)}[N_\omega (k\xi ,k\xi _2)]}{%
I'_\omega (k\xi _2)N'_\omega (k\xi _1,k\xi
_2)}-\frac{F^{(i)}[K_\omega (k\xi )]}{K'_\omega (k\xi _1)}\right]
 \nonumber
\end{equation}
is the 'interference' term. An equivalent formula for $\Delta
\langle T_{i}^{k}\rangle _N$ is obtained from Eq.
(\ref{intterm1N}) by replacements (\ref{replace}).

'Interference' term (\ref{intterm1N}) is finite for all $\xi
_1\leq \xi \leq\xi _2$, $\xi _1<\xi _2$, and diverges in the limit
$\xi _1\to \xi _2$. In this limit the main contribution into the
$\omega $--integral comes from the large values $\omega $.
Introducing a new integration variable $x=k\xi _1/\omega $ and
using the uniform asymptotic expansions for the Bessel modified
functions in the leading order one obtains that the quantities
$\Delta \langle T_{i}^{k}\rangle _N$ coincide with the VEV's for
two parallel plates in $d+1$--dimensional Minkowski spacetime with
separation $\xi _2-\xi _1$ \cite{Amb83,RomSah}. The corresponding
expressions are given by formulae (\ref{D2Mink0}),(\ref{D2Mink1})
with the opposite sign of the integral term on the right of
formula (\ref{D2Mink0}).

Now we turn to the Neumann vacuum effective pressures determining
the forces acting on the plate due to the presence of the second
plate (interaction forces). This force acting per unit surface of
the plate $\xi =\xi _2$, $p_{N{\mathrm{(int)}}}^{(2)}$ is defined
by the ${}_{1}^{1}$--component of the second term on the right of
formula (\ref{Tikbound1N}) at $\xi =\xi _2 $. The nonzero
contribution comes from the last term on the right of
Eq.(\ref{f1}) (with replacement (\ref{Ffunc})). Using the standard
Wronskian relation for the Bessel modified functions one obtains
\begin{equation}\label{pN2int}
  p_{N{\mathrm{(int)}}}^{(2)}(\xi _1,\xi _2)=\frac{A_d}{2\xi _2^2}\int _{0}^{\infty }
  dk k^{d-2}\int _{0}^{\infty }
  d\omega \frac{I'_{\omega }(k\xi _1)(1+\omega ^2/k^2\xi _2^2)}{I'_{\omega }
  (k\xi _2)N'_{\omega }(k\xi _1,k\xi _2)}.
\end{equation}
By a similar way for the interaction force per unit surface of the
first plate from the second term on the right of Eq.
(\ref{EMTNform2}) at $\xi =\xi _1$ we receive
\begin{equation}\label{pN1int}
  p_{N{\mathrm{(int)}}}^{(1)}(\xi _1,\xi _2)=\frac{A_d}{2\xi _1^2}\int _{0}^{\infty }
  dk k^{d-2}\int _{0}^{\infty }
  d\omega \frac{K'_{\omega }(k\xi _2)(1+\omega ^2/k^2\xi _1^2)}{K'_{\omega }
  (k\xi _1)N'_{\omega }(k\xi _1,k\xi _2)}.
\end{equation}
Note that pressures (\ref{pN2int}),(\ref{pN1int}) are independent
on the curvature coupling parameter. It can be seen that the
function $I'_{\omega }(z)/K'_{\omega }(z)$ is monotonic
decreasing, and as a result $ N'_{\omega }(k\xi _1,k\xi _2)<0$ for
$\xi _1<\xi _2$. In combination with Eqs. (\ref{pN2int}),
(\ref{pN1int}) it follows from here that
$p_{N{\mathrm{(int)}}}^{(i)}<0$, $i=1,2$, and hence, as in the
Dirichlet case, the Neumann interaction forces are always
attractive. By using that the function $z^4I'_{\omega
}(z)K'_{\omega }(z)/(z^2+\omega ^2)$ is monotonic decreasing (this
can be proved by using inequalities (\ref{Iineq1}),(\ref{Kineq2})
) we see that
\begin{equation}\label{ineqNp12}
  p_{N{\mathrm{(int)}}}^{(2)}(\xi _1,\xi _2)>p_{N{\mathrm{(int)}}}^{(1)}
  (\xi _1,\xi _2).
\end{equation}
In the limit $\xi _1\to \xi _2$ replacing the Bessel modified
functions by their uniform asymptotic expansions we can see that
to the leading order from Eqs. (\ref{pN2int}),(\ref{pN1int}) the
standard Casimir interaction force is obtained for two parallel
plates with separation $\xi _2-\xi _1$ in the $d+1$--dimensional
Minkowski spacetime.

\begin{figure}[tbph]
\begin{center}
\epsfig{figure=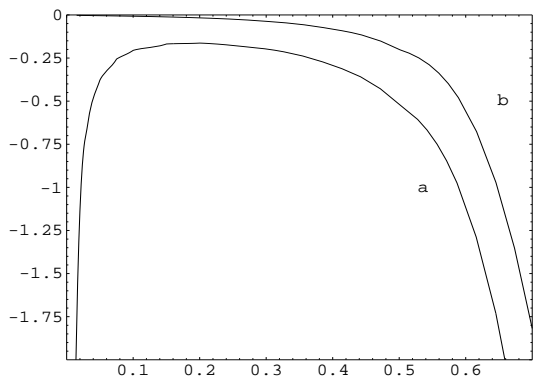,width=7cm,height=5cm}
\end{center}
\caption{ The $d=3$ vacuum effective pressures determining the
interaction forces per unit surface between Neumann parallel
plates, multiplied by $\xi ^{4}_2$, $\xi ^{4}_2 p_{D{\rm
(int)}}^{(1)}$ (curve a) and $\xi ^{4}_2 p_{D{\rm (int)}}^{(2)}$
(curve b) as functions of the ratio $\xi _1/\xi _2 $.}
\label{fig3pintN}
\end{figure}

From formulae (\ref{pN2int}), (\ref{pN1int}) one has
\begin{equation}
\xi _{1}\frac{\partial p_{N{\rm (int)}}^{(1)}}{\partial \xi _{2}}=-\xi _{2}%
\frac{\partial p_{N{\rm (int)}}^{(2)}}{\partial \xi
_{1}}=\frac{A_{d}}{2\xi
_{1}\xi _{2}}\int_{0}^{\infty }dkk^{D-2}\int_{0}^{\infty }d\omega \frac{%
(1+\omega ^{2}/k^{2}\xi _{1}^{2})(1+\omega ^{2}/k^{2}\xi _{2}^{2})}{%
N_{\omega }^{\prime 2}(k\xi _{1},k\xi _{2})}.  \label{pNxi2}
\end{equation}
As seen from here for a fixed value of $\xi _{2}$ ($\xi _{1}$) the
modulus of the interaction force acting on the plate at $\xi =\xi
_{2}$ ($\xi _{1}$) is a monotonic increasing (decreasing) function
on $\xi _{1}$ ($\xi _{2}$). For the other partial derivatives,
similar to the Dirichlet case, one has the relation (\ref{pDxi22})
with replacement $D\rightarrow N$. In particular, we can see that
$\partial p_{N{\rm (int)}}^{(2)}/\partial \xi _2>0$. The Neumann
effective pressures determining the interaction forces per unit
surface given by Eqs. (\ref{pN2int}), (\ref{pN1int}) are plotted
in Fig. \ref{fig3pintN} as functions of $\xi _1/\xi _2$ for the
case $d=3$.
As seen from Fig. \ref{fig2pintD} and Fig. \ref{fig3pintN} the
Dirichlet and Neumann vacuum interaction forces are numerically
close to each other. This is a consequence of that the
subintegrands in formulae (\ref{pint2}) and (\ref{pN2int}) and in
formulae (\ref{pint1}) and (\ref{pN1int}) are numerically close.
This can be also seen analytically by using relations
(\ref{Iineq1}),(\ref{Kineq2}).

\section{Electromagnetic field} \label{sec:elmag}

We now turn to the case of the electromagnetic field in the region
$\xi _{1}<\xi <\xi _{2}$. We will assume that the mirrors are
perfect conductors with the standard boundary conditions of
vanishing of the normal component of the magnetic field and the
tangential components of the electric field, evaluated at the
local inertial frame in which the conductors are instantaneously
at rest. By considerations similar to those given in Ref.
\cite{Candelas} for $d=3$, it can be seen that the corresponding
eigenfunctions for the vector potential $A^{\mu }$ may be resolved
into one transverse magnetic (TM) and $d-2$ transverse electric
(TE) (with respect to $\xi $-direction) modes $A^{\mu }_{\sigma
\alpha }$, $\sigma =0,1,\ldots ,d-2$, $\alpha
=({\mathbf{k}},\omega)$:
\begin{eqnarray}
A^{\mu }_{1 \alpha } & = & \left( -\xi \partial /\partial \xi
,-i\omega /\xi ,0,\ldots 0\right) \varphi _{0\alpha },\quad
\sigma =1, \quad {\mathrm{TM \, \, mode}}, \label{apot0}\\
A^{\mu }_{\sigma \alpha } & = & \epsilon ^{\mu }_{\sigma }\varphi
_{\sigma \alpha }
 ,\quad \sigma =0,2,\ldots ,d-2, \quad {\mathrm{TE \,\, modes}},
\label{apot1}
\end{eqnarray}
where the polarization vectors $\epsilon ^{\mu }_{\sigma }$ obey
the following relations
\begin{equation}\label{polariz}
\epsilon ^{0}_{\sigma }= \epsilon ^{1}_{\sigma }=0,\quad \epsilon
_{\sigma \mu } \epsilon ^{\mu }_{\sigma '} =-k^2\delta _{\sigma
\sigma '},\quad \epsilon ^{\mu }_{\sigma }k_{\mu }=0.
\end{equation}
From the perfect conductor boundary conditions one has the
following conditions for the scalar fields $\varphi _{\sigma
\alpha }$:
\begin{equation}
\varphi _{\sigma \alpha }| _{\xi =\xi _{1}}=\varphi _{\sigma
\alpha }| _{\xi =\xi _{2}}=0,\quad \sigma
=0,2,\ldots ,d-2, \quad \frac{\partial \varphi _{1\alpha }}{%
\partial \xi }| _{\xi =\xi _{1}}=\frac{\partial \varphi _{1\alpha }}{\partial \xi }|
_{\xi =\xi _{2}}=0.\label{elecbound}
\end{equation}
As a result the TE/TM modes correspond to the Dirichlet/Neumann
scalars. In the corresponding expressions for the eigenfunctions
$A^{\mu }_{\sigma \alpha }$ the normalization coefficient is
determined from the orthonormality relation
\begin{equation}
\int d{\bf x}\int_{\xi _{1}}^{\xi _{2}}\frac{d\xi }{\xi }A_{\sigma
\alpha }^{\mu }A_{\sigma '\alpha ^{\prime }\mu }^{\ast
}=-\frac{2\pi }{\omega }\delta _{\alpha \alpha ^{\prime }}\delta
_{\sigma \sigma ^{\prime }} . \label{elecnorm}
\end{equation}
On the base of this normalization condition for the separate
scalar modes one has
\begin{equation}
\varphi _{\sigma \alpha } =\frac{2\pi ^{1/2}}{k}C_ZZ_{i\omega
_{Zn}}(k\xi ,k\xi _2)\exp \left[ i\left( {\bf kx}-\omega _{Zn}\tau
\right) \right] , \label{fi0sol}
\end{equation}
where $Z=D$ for $\sigma =0,2,\ldots ,d-2$ and $Z=N$ for $\sigma =1
$, and the coefficients $C_D$ and $C_N$ are defined in accordance
with Eqs. (\ref{Dnormc}),(\ref{Nnormc}). Substituting the
eigenfunctions (\ref{apot0}), (\ref{apot1}) into the mode sum
formula
\begin{equation}
\langle 0| T_{i}^{k}| 0\rangle =\sum_{\sigma =0}^{d-2}\int d{\bf k}%
\sum_{\omega _{Zn} }T_{i}^{k}\{A_{\sigma \alpha \mu },A_{\sigma
\alpha \mu }^{\ast }\}, \label{eleksum}
\end{equation}
with the standard bilinear form for the electromagnetic field EMT
one finds
\begin{equation}\label{elvevEMT}
  \langle 0| T_{i}^{k}| 0\rangle =\delta _i^k \frac{\pi ^{(d-1)/2}}{
  \Gamma \left( \frac{d-1}{2}\right)}\int _{0}^{\infty }
  dk\, k^d\sum_{\sigma =0,1}\beta _{\sigma }\sum _{n=1}^{\infty }
  C_Z^2f_{{\mathrm{em}}}^{(i)}[Z_{i
  \omega _{Zn}}(k\xi ,k\xi _{2})], \quad \beta_0=d-2,\quad \beta
  _1=1,
\end{equation}
where $\beta _0$ and $\beta _1$ are the numbers of the independent
polarization states for TE and TM modes respectively. In Eq.
(\ref{elvevEMT}) for a given function $G(z)$ the following
notations are introduced
\begin{eqnarray}
f_{{\mathrm{em}}}^{(0)}[G(z)] &=&\left| \frac{dG(z)}{dz}\right|
^{2}+\left( 1+\frac{
\omega ^{2}}{z^{2}}\right) |G(z)|^{2},  \nonumber \\
f_{{\mathrm{em}}}^{(1)}[G(z)] &=&-\left| \frac{dG(z)}{dz}\right|
^{2}+\left( 1-\frac{
\omega ^{2}}{z^{2}}\right) |G(z)|^{2},  \label{femfunk} \\
f_{{\mathrm{em}}}^{(i)}[G(z)] &=&
\frac{d-5}{d-1}|G(z)|^{2}+\frac{d-3}{d-1}\left[ \left|
\frac{dG(z)}{dz}\right| ^{2}-\frac{\omega ^{2}}{z^{2}}|G(z)|^{2}
\right] ,\quad i=2,3,\ldots ,d.\nonumber
\end{eqnarray}
By making use of the summation formulae derived in the Appendix A
the VEV's are presented in the form
\begin{equation}
\langle 0| T_{i}^{k}| 0\rangle =\delta _{i}^{k} \frac{A_d}{2}%
\int_{0}^{\infty }dkk^{d}\int_{0}^{\infty }d\omega \sum_{\sigma
=0,1} \beta _{\sigma }\left\{ \frac{\sinh \pi \omega }{\pi }
\,f_{{\mathrm{em}}}^{(i)}[\tilde{Z}_{i\omega }(k\xi ,k\xi _{2})]-
\frac{I_{\omega }^{(\sigma )}(k\xi
_{1})F_{{\mathrm{em}}}^{(i)}[Z_{\omega }(k\xi ,k\xi
_{2})]}{I_{\omega }^{(\sigma )}(k\xi _{2})Z_{\omega }^{(\sigma
)}(k\xi _1,k\xi _2)}\right\}, \label{EMTelec}
\end{equation}
where $I_{\omega }^{(0)}=I_{\omega}$, $I_{\omega
}^{(1)}=I'_{\omega }$, and the same notations for the functions
$K_{\omega }$, $Z_{\omega }$. The functions $F_{em}^{(i)}$ are
obtained from Eqs. (\ref{femfunk}) replacing $\omega \rightarrow
i\omega $:
\begin{equation}
F_{{\mathrm{em}}}^{(i)}[G(z)]=f_{{\mathrm{em}}}^{(i)}[G(z),\omega
\rightarrow i\omega ]. \label{Felecfunc}
\end{equation}
It can be easily checked that the components (\ref{EMTelec}) obey
the covariant conservation equation and the corresponding EMT is
traceless for $d=3$. The first term in the figure braces of Eq.
(\ref{EMTelec}) corresponds to the VEV induced by a single plate
at $\xi =\xi _2$ in the region $\xi <\xi _2$. For the case $d=3$
they are investigated in Ref. \cite{Saharian1}. The generalization
for an arbitrary $d$ is straightforward and these quantities are
presented in the form
\begin{equation}\label{el1b}
  \langle 0|T_i^k|0\rangle ^{(1b)}(\xi _2,\xi )=\langle
  0_R|T_i^k|0_R\rangle -\frac{1}{2}\delta _{i}^{k}A_d\int _{0}^{\infty }
  dkk^d\int _{0}^{\infty }d\omega \sum _{\sigma =0,1} \beta _\sigma
  \frac{K_{\omega }^{(\sigma)}(k\xi _2)}{I_{\omega }^{(\sigma)}(k\xi _2)}
  F_{{\mathrm{em}}}^{(i)}[I_{\omega }(k\xi )],
\end{equation}
where $\langle  0_R|T_i^k|0_R\rangle $ are the VEV's for the
Fulling--Rindler electromagnetic vacuum without boundaries. By the
way similar to that given in Ref. \cite{Saharian1} for the case of
a scalar field, it can be seen that
\begin{equation}\label{elfulrin}
  \langle 0_R|T_i^k|0_R\rangle = \langle 0_M|T_i^k|0_M\rangle -
  \frac{\delta _{i}^{k}(d-1)\xi ^{-d-1}}{2^{d-1}\pi ^{d/2}\Gamma (d/2)}%
\int_{0}^{\infty }\frac{\omega ^df_{0{\mathrm{em}}}^{(i)}(\omega
)d\omega }{e^{2\pi \omega }+(-1)^d} \prod _{l=1}^{l_m}\left[
\left( \frac{d-1-2l}{2\omega } \right) ^2+1 \right] ,
\end{equation}
where $l_{m}=d/2-1$ for even $d>2$ and $l_{m}=(d-1)/2$ for odd
$d>1$, and the value for the product over $l$ is equal to 1 for
$d=1,2,3$. In Eq. (\ref{elfulrin}) we have introduced notations
\begin{eqnarray}
f_{0{\mathrm{em}}}^{(0)}(\omega )
&=&-df_{0{\mathrm{em}}}^{(1)}(\omega )=1+\frac{(d-1)^2}{4\omega
^{2}}, \label{f00el}
\\
f_{0{\mathrm{em}}}^{(i)}(\omega )& =&
f_{0{\mathrm{em}}}^{(1)}(\omega )+\frac{(d-1)(d-3)}{4\omega ^2 },
\quad i=2,\ldots ,d.  \nonumber
\end{eqnarray}
For physically most important case $d=3$, formula (\ref{elfulrin})
leads to the standard result derived by Candelas and Deutsch in
Ref. \cite{Candelas}.

An alternative form for the vacuum EMT in the region between two
plates is
\begin{eqnarray}\label{eleqform}
  \langle 0|T_i^k|0\rangle & = & \langle
  0_R|T_i^k|0_R\rangle +\langle 0|T_i^k|0\rangle ^{(1b)}(\xi _1,\xi
  )- \nonumber \\
  &-& \frac{1}{2}\delta _{i}^{k}A_d\int _{0}^{\infty }
  dkk^d\int _{0}^{\infty }d\omega \sum _{\sigma =0,1}\beta _\sigma
  \frac{K_{\omega }^{(\sigma )}(k\xi
_{2})F_{{\mathrm{em}}}^{(i)}[Z_{\omega }(k\xi ,k\xi
_{1})]}{K_{\omega }^{(\sigma )}(k\xi _{1})Z_{\omega }^{(\sigma
)}(k\xi _1,k\xi _2)},
\end{eqnarray}
where $\langle 0|T_i^k|0\rangle ^{(1b)}(\xi _1,\xi )$ is the
vacuum EMT induced by a single boundary at $\xi =\xi _1$ in the
region $\xi >\xi _1$. The latter is obtained from (\ref{el1b}) by
replacements (\ref{replace}). For the interaction force
$p^{(i)}_{{\mathrm{em(int)}}}$, $i=1,2$ per unit area of the plate
at $\xi =\xi _i$ from Eqs. (\ref{EMTelec}) and (\ref{eleqform})
one obtains
\begin{eqnarray}\label{pem1}
  p^{(1)}_{{\mathrm{em(int)}}}&=&-\frac{A_d}{2\xi _1^2}\int _{0}^{\infty
  }dkk^{d-2}\int_{0}^{\infty }d\omega \sum _{\sigma =0,1}(-1)^{\sigma }\beta _\sigma
  \frac{K_{\omega }^{(\sigma )}(k\xi
_{2})}{K_{\omega }^{(\sigma )}(k\xi _{1})}\frac{(1+\omega
^2/k^2\xi _1^2)^{\sigma }}{Z_{\omega }^{(\sigma )}(k\xi _1,k\xi
_2)} , \\
  p^{(2)}_{{\mathrm{em(int)}}}&=&-\frac{A_d}{2\xi _2^2}\int _{0}^{\infty
  }dkk^{d-2}\int_{0}^{\infty }d\omega \sum _{\sigma =0,1}(-1)^{\sigma }\beta _\sigma
  \frac{I_{\omega }^{(\sigma )}(k\xi
_{1})}{I_{\omega }^{(\sigma )}(k\xi _{2})}\frac{(1+\omega
^2/k^2\xi _2^2)^{\sigma }}{Z_{\omega }^{(\sigma )}(k\xi _1,k\xi
_2)} .\label{pem2}
\end{eqnarray}
Recalling that $(-1)^\sigma Z_{\omega }^{(\sigma )}>0$ we see the
electromagnetic interaction forces are attractive. Note that
$p^{(i)}_{{\mathrm{em(int)}}}=(d-2)p^{(i)}_{D{\mathrm{(int)}}}+
p^{(i)}_{N{\mathrm{(int)}}}$. In the limit $\xi _1\to \xi _2$ and
to the leading order over $1/(\xi _2-\xi _1)$ from these
expressions the electromagnetic Casimir interaction force between
plates in the Minkowski spacetime is obtained.

\section{Conclusion} \label{sec:conc}

It is well known that the uniqueness of vacuum state is lost when
we work within the framework of quantum field theory in a general
curved spacetime or in non--inertial frames. In this paper we have
considered vacuum expectation values of the energy-momentum tensor
for scalar and electromagnetic fields between two infinite
parallel plates moving by uniform proper acceleration, assuming
that the fields are prepared in the Fulling-Rindler vacuum state.
As the boundaries are static in the Rindler coordinates no Rindler
quanta are created and the only effect of the imposition of
boundary conditions on quantum fields is the vacuum polarization.
For the scalar case the both Dirichlet and Neumann boundary
conditions are investigated. The VEV's are presented in the form
of mode sums involving series over zeros $\omega =\omega _{Dn}$ or
$\omega =\omega _{Nn}$ of the functions $D_{i\omega }(k\xi _1,k\xi
_2)$ and $N'_{i\omega }(k\xi _1,k\xi _2)$ respectively. To sum
these series we derive in Appendix \ref{section:App1} summation
formulae for these types of series using the generalized
Abel-Plana formula. The application of these formulae allows to
extract from the VEV's the parts due to the single plate. The
latters are investigated previously in Refs.
\cite{Candelas,Saharian1}. The boundary induced parts are
presented in two alternative forms, Eqs. (\ref{Tikbound1}),
(\ref{EMTDform2}), for the Dirichlet case, and Eqs.
(\ref{TikboundN}),(\ref{Tikbound1N}) for the Neumann case. Various
limiting cases are studied. In particular, in the limit when the
left plate coincides with the Rindler horizon the corresponding
VEV's are the same as for a single plate geometry. The vacuum
forces acting on boundaries contain two terms. The first ones are
the forces acting on a single boundary then the second boundary is
absent. Due to the well--known surface divergences in the VEV's of
the energy-momentum tensor these forces are infinite and need an
additional regularization. The another terms in the vacuum forces
are finite and are induced by the presence of the second boundary
and correspond to the interaction forces between the plates. These
forces per unit surface do not depend on the curvature coupling
parameter $\zeta $ and are determined by formulae
(\ref{pDxi2}),(\ref{pDxi22}) for the Dirichlet scalar and by
formulae (\ref{pN2int}),(\ref{pN1int}) for the Neumann scalar, and
are always attractive for both plates. In particular, they are the
same for conformally and minimally coupled scalars. For given $\xi
_1$, $\xi _2 $ the modulus of the interaction force is larger for
the plate at $\xi =\xi _1$ (see inequalities (\ref{pint12ineq})
and (\ref{ineqNp12})). For small distances between the plates at
the leading order the standard Casimir result on background of the
Minkowski vacuum is rederived. The case of the electromagnetic
field is considered with the perfect conductor boundary conditions
in the local inertial frame in which the boundaries are
instantaneously at rest. The corresponding eigenmodes are
superposition of TE and TM modes with Dirichlet and Neumann
boundary conditions respectively. The VEV's of the electromagnetic
EMT in the region between the plates are given by formulae
(\ref{EMTelec}) and (\ref{eleqform}). The corresponding vacuum
interaction forces per unit surface, Eqs.
(\ref{pem1}),(\ref{pem2}), are obtained by summing the Dirichlet
and Neumann $d=3$ scalar forces, and are attractive for all values
of the proper accelerations for the plates.

The results obtained in this paper can be applied to the geometry
of two parallel plates near the 'Rindler wall'. With the $x$
coordinate perpendicular to the wall and with the $(x^2,x^3)$
plane located at the centre of the wall, $x=0$, the static
plane--symmetric line element can be written as
\begin{equation}\label{metricwall1}
  ds^2=e^{\nu (x)}dt^2-dx^2-e^{\lambda (x)}d{\mathbf{x}}^2,\quad
{\mathbf{x}}=(x^2,x^3),
\end{equation}
where $\nu (x)$ and $\lambda (x)$ are even functions. For this
metric the Einstein equations with the diagonal matter
energy-momentum tensor $T_i^{(m)k}={\mathrm{diag}}(\varepsilon
^{(m)},-p^{(m)},-p^{(m)},-p^{(m)})$ admit two classes of
solutions. For the first one $\lambda '(0)>0$, and the
corresponding external solution (the solution in the region
$x>x_s$, where $T_{ik}^{(m)}=0$, with $x=x_s$ being the boundary
of the wall) is described by the standard Taub metric
\cite{Taub51}. For the second class of internal solutions $\lambda
'(0)<0 $, and the external solution is presented by the metric
\begin{equation}\label{metricRwall}
  ds^2_{{\mathrm{ext}}}=e^{\nu _s}[1+2\pi \sigma _s(x-x_s)]^2dt^2-
  dx^2-e^{\lambda _s}d{\mathbf{x}}^2,
\end{equation}
where $\nu _s=\nu (x_s)$, $\lambda _s=\lambda (x_s)$, and
\begin{equation}\label{sigmas}
  \sigma _s=2e^{-\nu _s/2-\lambda _s}\int _{0}^{x_s}
  \left( \varepsilon ^{(m)}+3p^{(m)}\right) e^{\nu /2+\lambda }dx
\end{equation}
is the mass per unit surface of the wall. For a given equation of
state $p^{(m)}=p^{(m)}(\varepsilon ^{(m)})$ the parameters $\nu
_s, \lambda _s,x_s$ are functions of the central pressure
$p^{(m)}|_{x=0}$, and are determined by the internal solution of
the Einstein equations (see Ref. \cite{Avak01} for the case of the
equation of state corresponding to the incompressible liquid). Now
redefining
\begin{equation}\label{walltoRind}
  \xi (x)=x-x_s+\frac{1}{2\pi \sigma _s},\quad \tau =2\pi \sigma
  _s e^{\nu _s/2}t,\quad e^{\lambda _s/2}x^i\to x^i,\quad i=2,3,
\end{equation}
from Eq. (\ref{metricRwall}) we obtain the Rindler metric in the
form (\ref{metric}). Hence, the VEV's for the EMT in the region
between two plates located at $x=x_1$ and $x=x_2$, $x_i>x_s$ near
the 'Rindler wall' are obtained from the results given above
substituting $\xi _i=\xi (x_i)$, $i=1,2$ and $\xi =\xi (x)$. Note
that for $\sigma _s>0$, $x\geq x_s$ one has $\xi (x) \geq \xi
(x_s)>0$ and the Rindler metric is regular everywhere.

\section*{Acknowledgements}

We are grateful to Professor E. Chubaryan and Professor A.
Mkrtchyan for general encouredgement and suggestions, and to L.
Grigoryan and R. Davtyan for useful discussions. This work was
supported by the Armenian National Science and Education Fund
(ANSEF) Grant No. PS14-00 and by the Armenian Ministry of
Education and Science Grant No. 0887.

\appendix

\section{Summation formulae over zeros of $D_{iz}$ and $N_{iz}^{\prime }$}
\label{section:App1}

In this section we will derive a summation formula over zeros
$z=\omega _{Dk}$ of the function
\begin{equation}
D_{iz}(x,y)=I_{iz}(y)K_{iz}(x)-I_{iz}(x)K_{iz}(y),\quad y>x.  \label{Dfunc}
\end{equation}
As we saw in section \ref{sec:Dir} the VEV's of the EMT for the
Dirichlet scalar between two plates in the Fulling-Rindler vacuum
are expressed in the form of series over these zeros. To derive a
summation formula we use the generalized Abel-Plana formula
\cite{Sahrev}. Let us choose in this formula
\begin{eqnarray}
f(z) &=&\frac{2i}{\pi }\sinh \pi z\,F(z),  \label{DfgtoAP} \\
g(z) &=&\frac{I_{iz}(y)I_{-iz}(x)+I_{iz}(x)I_{-iz}(y)}{D_{iz}(x,y)}F(z),
\nonumber
\end{eqnarray}
with a meromorphic function $F(z)$ having poles $z=z_{k}$ in the
right half-plane ${\rm Re}\,z\geq 0$. The sum and difference of
functions (\ref {DfgtoAP}) are presented in the form
\begin{equation}
g(z)\pm f(z)=\frac{2I_{\mp iz}(x)I_{\pm iz}(y)}{D_{iz}(x,y)}F(z).
\label{Dgpmf}
\end{equation}
By taking into account that the zeros $\omega _{Dk}$ are simple
poles of the function $g(z)$ for the function $R[f(z),g(z)]$ in
the generalized Abel-Plana formula one obtains
\begin{eqnarray}
R[f(z),g(z)] &=&2\pi i\left[ \sum_{k=1}^{\infty }\frac{I_{-iz}(y)I_{iz}(x)}{%
\frac{\partial }{\partial z}D_{iz}(x,y)}F(z)|_{z=\omega
_{Dk}}+\right.
\label{DRfg} \\
&+&\left. \sum_{k}{\rm Res}_{z=z_{k}}\frac{F(z)}{D_{iz}(x,y)}I_{i{\rm sgn}(%
{\rm Im}z_{k})z}(y)I_{-i{\rm sgn}({\rm Im}z_{k})z}(x)\right] ,
\nonumber
\end{eqnarray}
where the zeros $\omega _{Dk}$ are arranged in ascending order. As
a result we obtain the following summation formula
\begin{eqnarray}
\sum_{k=1}^{\infty }\frac{I_{-iz}(y)I_{iz}(x)}{\frac{\partial }{\partial z}%
D_{iz}(x,y)}F(z)|_{z=\omega _{Dk}} &=&\frac{1}{\pi
^{2}}\int_{0}^{\infty
}\sinh \pi z\,F(z)dz-  \label{Dsumformula} \\
&-&\sum_{k}{\rm Res}_{z=z_{k}}\frac{F(z)}{D_{iz}(x,y)}I_{i{\rm sgn}({\rm Im%
}z_{k})z}(y)I_{-i{\rm sgn}({\rm Im}z_{k})z}(x)-  \nonumber \\
&-&\frac{1}{2\pi }\int_{0}^{\infty }dz\frac{F(ze^{\pi
i/2})+F(ze^{-\pi i/2}) }{
I_{-z}(y)K_{z}(x)-I_{-z}(x)K_{z}(y)}I_{z}(x)I_{-z}(y).  \nonumber
\end{eqnarray}
Here the condition for the function $F(z)$ is easily obtained from
the corresponding condition in the Generalized Abel--Plana formula
by using the asymptotic formulae for the Bessel modified function
and has the form
\begin{equation}
|F(z)|<\epsilon (|z|)e^{-\pi z}\left( \frac{y}{x}\right) ^{2|{\rm Im}%
z|},\quad {\rm Re\,}z>0,\quad |z|\rightarrow \infty ,  \label{condforAPF2pl}
\end{equation}
where $|z|\epsilon (|z|)\rightarrow 0$ when $|z|\rightarrow \infty $.

A similar formula can be obtained for the series over zeros $z=\omega _{Nk}$%
, $k=1,2,\ldots $ of the function
\begin{equation}
N_{iz}^{\prime }(x,y)=I_{iz}^{\prime }(y)K_{iz}^{\prime }(x)-K_{iz}^{\prime
}(y)I_{iz}^{\prime }(x),\quad \;y>x.  \label{Nfunc}
\end{equation}
For this let us substitute in the Generalized Abel-Plana formula \cite
{Sahrev}
\begin{eqnarray}
f(z) &=&\frac{2i}{\pi }\sinh \pi z\,F(z)  \label{NfgtoAP} \\
g(z) &=&\frac{I^{\prime }{}_{iz}(y)I^{\prime }{}_{-iz}(x)+I^{\prime
}{}_{iz}(x)I^{\prime }{}_{-iz}(y)}{N_{iz}^{\prime }(x,y)}F(z).  \nonumber
\end{eqnarray}
Using these expressions it can bee easily seen that
\begin{equation}
g(z)\pm f(z)=\frac{2I_{\mp iz}^{\prime }(x)I_{\pm iz}^{\prime }(y)}{%
N_{iz}^{\prime }(x,y)}F(z).  \label{Ngpmf}
\end{equation}
For the function $R[f(z),g(z)]$ now one obtains
\begin{eqnarray}
R[f(z),g(z)] &=&2\pi i\left[ \sum_{k=1}^{\infty }\frac{I^{\prime
}{}_{-iz}(y)I^{\prime }{}_{iz}(x)}{\frac{\partial }{\partial z}%
N_{iz}^{\prime }(x,y)}F(z)|_{z=\omega _{Nk}}+\right.  \label{NRfg} \\
&+&\left. \sum_{k}{\rm Res}_{z=z_{k}}\frac{F(z)}{N_{iz}^{\prime }(x,y)}I_{i%
{\rm sgn}({\rm Im}z_{k})z}(y)I_{-i{\rm sgn}({\rm
Im}z_{k})z}(x)\right] . \nonumber
\end{eqnarray}
As a result we obtain the following summation formula
\begin{eqnarray}
\sum_{k=1}^{\infty }\frac{I^{\prime }{}_{-iz}(y)I^{\prime }{}_{iz}(x)}{\frac{%
\partial }{\partial z}N_{iz}^{\prime }(x,y)}F(z)|_{z=\omega _{Nk}} &=&%
\frac{1}{\pi ^{2}}\int_{0}^{\infty }\sinh \pi z\,F(z)dz-  \label{Nsumformula}
\\
&-&\sum_{k}{\rm Res}_{z=z_{k}}\frac{F(z)}{N_{iz}^{\prime }(x,y)}I_{i{\rm %
sgn}({\rm Im}z_{k})z}(y)I_{-i{\rm sgn}({\rm Im}z_{k})z}(x)-  \nonumber \\
&-&\frac{1}{2\pi }\int_{0}^{\infty }dz\frac{F(ze^{\pi
i/2})+F(ze^{-\pi i/2})}{I_{-z}^{\prime }(y)K_{z}^{\prime
}(x)-I_{-z}^{\prime }(x)K_{z}^{\prime }(y)}I_{z}^{\prime
}(x)I_{-z}^{\prime }(y),  \nonumber
\end{eqnarray}
where the corresponding condition for the function $F(z)$ has the
form (\ref{condforAPF2pl}).

\section{$d=1$ case: Direct evoluation} \label{section:App2}

For $d=1$ case the linearly independent solutions to equation (%
\ref{fiequ}) are $e^{\pm i\omega \ln \xi }$. The normalized
eigenfunctions satisfying Dirichlet boundary conditions
(\ref{Dboundcond}) are in form
\begin{equation}
\varphi _n^D=\frac{e^{-i\omega \tau }}{\sqrt{\pi n}}%
\sin (\alpha n), \quad \omega =\frac{\pi n}{\ln (\xi _{2}/\xi
_1)}, \quad n=1,2,\ldots, \label{D1eigenfunc}
\end{equation}
where we use the notation
\begin{equation}\label{alfnot}
  \alpha =\frac{\pi \ln (\xi _2/\xi )}{\ln (\xi _2/\xi _1)}.
\end{equation}
Substituting eigenfunctions (\ref{D1eigenfunc}) into mode--sum
formula (\ref{EMTvev}) and applying to the sum over $n$ the
Abel--Plana summation formula one finds
\begin{eqnarray}
  \langle 0_D|T_{i}^{k}|0_D\rangle -\langle 0_M|T_{i}^{k}|0_M\rangle
  &=& \langle T_{i}^{k}\rangle _{{\mathrm{(sub)}}}^{(R)}+
  \langle T_{i}^{k}\rangle ^{(1b)}_D(\xi _2,\xi )+
  \zeta\frac{\alpha ^2/\sin ^2\alpha -1}{2\pi \xi ^2\ln ^2(\xi _2/\xi)}
{\mathrm{diag}}(1,0) -\nonumber \\
  &-& \frac{1}{2\pi \xi ^2}\left[ \frac{\pi ^2}{12\ln ^2(\xi _2/\xi _1)}+
  \frac{\zeta }{\ln (\xi _2/\xi )}(\alpha \cot \alpha -1)\right]
  {\mathrm{diag}}(1,-1). \label{D1vevEMT}
\end{eqnarray}
Here the subtracted purely Fulling--Rindler part without
boundaries, $\langle T_{i}^{k}\rangle _{{\mathrm{(sub)}}}^{(R)}$,
and the part induced by a single boundary at $\xi =\xi _2$ are
given by formulae \cite{Saharian1}
\begin{equation}
   \langle T_{i}^{k}\rangle _{{\mathrm{(sub)}}}^{(R)}=
  \frac{1}{2\pi \xi ^2}\left( \zeta -\frac{1}{12}\right)
  {\mathrm{diag}}(1,-1), \label{FR1D}
\end{equation}
\begin{equation}
   \langle T_{i}^{k}\rangle ^{(1b)}_D(\xi _2,\xi )=
  \frac{\zeta }{2\pi \xi ^2\ln (\xi /\xi _2)}{\mathrm{diag}}(1+
  1/\ln(\xi /\xi _2),-1). \label{b11D}
\end{equation}
Note that the expression (\ref{b11D}) for a single boundary part
is valid for both regions $\xi <\xi _2$ and $\xi >\xi _2$. Now for
the vacuum interaction forces between the plates one obtains
\begin{equation}\label{pDd1}
  p_{D{\mathrm{(int)}}}^{(i)}=-\frac{\pi }{24\xi _i^2\ln ^2(\xi _2/\xi _1)},\quad
  i=1,2.
\end{equation}
In the limit $\xi _1\to \xi _2$ to the leading order we recover
the standard Casimir result on background of the 2D Minkowski
spacetime.

For the case of the Neumann boundary conditions (\ref{Nboundcond})
the normalized eigenfunctions have the form
\begin{equation}\label{N1eigenfunc}
  \varphi _n^N=\frac{e^{-i\omega \tau }}{\sqrt{\pi n}}%
\cos (\alpha n), \quad n=0,1,2,\ldots,
\end{equation}
where $\omega $ and $\alpha $ are given by the same relations
(\ref{D1eigenfunc}) and (\ref{alfnot}) as in the Dirichlet case.
The substitution of these eigenfunctions into the mode--sum
formula shows that the VEV's of the EMT for the Neumann boundary
conditions can be obtained from the corresponding formula for the
Dirichlet case, Eq. (\ref{D1vevEMT}), replacing in the boundary
part $\zeta \to -\zeta $.

\section{Alternative representation for the VEV's}\label{section:App3}

As a solution to equation (\ref{fiequ}) satisfying first boundary condition (%
\ref{Dboundcond}) one could take the function
\begin{equation}
D_{i\omega }(k\xi ,k\xi _{1})=I_{i\omega }(k\xi _{1})K_{i\omega }(k\xi
)-K_{i\omega }(k\xi _{1})I_{i\omega }(k\xi ).  \label{D1}
\end{equation}
Now from the boundary condition on the plate $\xi =\xi _{2}$ (\ref
{Dboundcond}) we find the possible values for $\omega $ being
roots to the equation (\ref{Deigfreq}). For the normalization
coefficient we receive
\begin{equation}
C_{D}^{2}=\frac{1}{\left( 2\pi \right) ^{d-1}}\frac{I_{i\omega }(k\xi _{2})}{%
I_{i\omega }(k\xi _{1})\frac{\partial D_{i\omega }(k\xi _{1},k\xi _{2})}{%
\partial \omega }}| _{\omega =\omega _{Dn}}.  \label{D1normc}
\end{equation}
The VEV's of the energy - momentum tensor are obtained in a
diagonal form
\begin{equation}
\langle 0_D| T_{i}^{k}| 0_D\rangle =\pi A_d\delta
_{i}^{k}\int_{0}^{\infty }dkk^{d}\sum_{n=1}^{\infty
}\frac{I_{i\omega }(k\xi _{2})}{I_{i\omega }(k\xi
_{1})\frac{\partial D_{i\omega }(k\xi _{1},k\xi _{2})}{\partial \omega }}%
f^{(i)}[D_{i\omega }(k\xi ,k\xi _{1})]| _{\omega =\omega _{Dn}}.
\label{EMT1diag}
\end{equation}
For the further evoluation of VEV's (\ref{EMT1diag}) we can apply
to the sum over $n$ summation formula (\ref{Dsumformula}). This
gives
\begin{eqnarray}
\langle 0_D| T_{i}^{k}| 0_D\rangle &=&A_d\delta
_{i}^{k}\int_{0}^{\infty }dkk^{d} \int_{0}^{\infty }d\omega
\,\left\{ \frac{\sinh \pi \omega }{\pi } f^{(i)}[
\tilde{D}_{i\omega }(k\xi ,k\xi _{1})]-\right.  \nonumber \\
&&-\left. \frac{I_{-\omega }(k\xi _{2})F^{(i)}[D_{\omega }(k\xi
,k\xi _{1})]}{I_{-\omega }(k\xi _{1})D_{\omega }(k\xi _{1},k\xi
_{2})}\right\} .  \label{EMT1diag1}
\end{eqnarray}
This form of the VEV's is equivalent to Eq. (\ref{EMTDdiag1}). To
see this let us consider the quantities
\begin{equation}
q_{j}^{(i)}=\frac{1}{\pi }\int_{0}^{\infty }d\omega \,\sinh \pi
\omega \,f^{(i)}[\tilde{D}_{i\omega }(k\xi ,k\xi
_{j})]-\int_{0}^{\infty }d\omega \frac{I_{\omega }(k\xi
_{1})I_{-\omega }(k \xi _{2})}{D_{\omega }(k\xi _{1},k\xi _{2})}
\frac{F^{(i)}[D_{\omega }(k\xi ,k\xi _{j})]}{I_{\omega }(k\xi
_{j})I_{-\omega }(k \xi _{j})},  \label{apqij}
\end{equation}
where $j=1,2$. Two representations (\ref{EMTDdiag1}) and
(\ref{EMT1diag1}) will be equivalent if
\begin{equation}
q_{1}^{(i)}=q_{2}^{(i)}.  \label{qequal}
\end{equation}
To prove this let us consider the difference
\begin{equation}
q_{2}^{(i)}-q_{1}^{(i)}=\frac{1}{\pi }\int_{0}^{\infty }d\omega
\,\sinh \pi \omega \,s^{(i)}-\int_{0}^{\infty }d\omega
\frac{I_{\omega }(k\xi _{1})I_{-\omega }(k \xi _{2})}{D_{\omega
}(k\xi _{1},k\xi _{2})}S^{(i)}, \label{equal12}
\end{equation}
where we have introduced the notations
\begin{eqnarray}
s^{(i)} &=&f^{(i)}[\tilde{D}_{i\omega }(k\xi ,k\xi _{2})]-f^{(i)}[%
\tilde{D}_{i\omega }(k\xi ,k\xi _{1})],  \label{siSi} \\
S^{(i)} &=& \sum _{j=1,2}(-1)^{j}\frac{F^{(i)}[D_{\omega }(k\xi
,k\xi _{j})]}{I_{\omega }(k\xi _{j})I_{-\omega }(k \xi _{j})}.
\nonumber
\end{eqnarray}
By using the standard relation between the Bessel modified
functions it can be seen that the first integral in formula
(\ref{equal12}) can be presented as
\begin{equation}
\frac{i}{2}\int_{0}^{\infty }\frac{I_{i\omega }(k\xi _{1})I_{-i\omega }(k\xi
_{2})-I_{-i\omega }(k\xi _{1})I_{i\omega }(k\xi _{2})\,}{I_{i\omega }(k\xi
_{2})K_{i\omega }(k\xi _{1})-I_{i\omega }(k\xi _{1})K_{i\omega }(k\xi _{2})}%
\,s^{(i)}d\omega ,
\end{equation}
where the function $s^{(i)}/(I_{i\omega }(k\xi _{2})K_{i\omega
}(k\xi _{1})-I_{i\omega }(k\xi _{1})K_{i\omega }(k\xi _{2}))$ has
no poles. \ For the term with the first (second) summand in the
numerator rotating the integration contour by angle $-\pi /2$
($\pi /2$) in $\omega $ complex plane and noting that the
integrals over arcs with large radius vanish (subintegrand behaves
as $(\xi /\xi _{2})^{2|{\rm Im}\omega |}$) we see that
\begin{equation}
\frac{i}{2}\int_{0}^{\infty }\frac{I_{i\omega }(k\xi _{1})I_{-i\omega }(k\xi
_{2})-I_{-i\omega }(k\xi _{1})I_{i\omega }(k\xi _{2})\,}{I_{i\omega }(k\xi
_{2})K_{i\omega }(k\xi _{1})-I_{i\omega }(k\xi _{1})K_{i\omega }(k\xi _{2})}%
\,s^{(i)}d\omega =\int_{0}^{\infty }d\omega \frac{I_{\omega }(k\xi
_{1})I_{-\omega }(k\xi _{2})}{D_{\omega }(k\xi _{1},k\xi
_{2})}S^{(i)}. \label{relap2}
\end{equation}
Hence, the difference (\ref{equal12}) is equal to zero, which
directly proves Eq. (\ref{qequal})

By taking into account Eq. (\ref{apqij}) from Eq. (\ref{qequal})
in the limit $\xi _{2}\rightarrow \infty $ one obtains the
following useful relation
\begin{eqnarray}
\frac{1}{\pi }\int_{0}^{\infty }d\omega \,\sinh \pi \omega \,f^{(i)}[%
\tilde{D}_{i\omega }(k\xi ,k\xi _{1})] &=&\frac{1}{\pi
}\int_{0}^{\infty }d\omega \,\sinh \pi \omega \,f^{(i)}[K_{i\omega
}(k\xi )]+  \label{limcond}
\\
&&+\int_{0}^{\infty }d\omega \left\{ \frac{F^{(i)}[D_{\omega }(k\xi ,k\xi
_{1})]}{I_{-\omega }(k\xi _{1})K_{\omega }(k\xi _{1})}-\frac{I_{\omega
}(k\xi _{1})}{K_{\omega }(k\xi _{1})}F^{(i)}[K_{\omega }(k\xi ]\right\}
\nonumber
\end{eqnarray}
Substituting Eq. (\ref{limcond}) into Eq. (\ref{EMT1diag1}) one
finds formula (\ref{EMTDform2}).


\begin{thebibliography}{99}

\bibitem{Casimir} H. B. G. Casimir, Proc. Kon. Nederl. Akad.
Wet. {\bf 51}, 793 (1948).

\bibitem{Mostepanenko} V. M. Mostepanenko and N. N. Trunov, {\it The
Casimir effect and its applications} (Oxford University Press,
Oxford, 1997).

\bibitem{Plunien} G. Plunien, B. Muller and W.Greiner, Phys. Rep.
{\bf 134}, 87 (1986).

\bibitem{Milton} K. A. Milton, {\it The Casimir Effect: Physical Manifestation
of Zero--Point Energy} (World Scientific, Singapore, 2002).

\bibitem{Lamor} S. K. Lamoreaux, Am. J. Phys. {\bf 67}, 850 (1999).

\bibitem{Bordag} M. Bordag (Ed.), {\it The Casimir Effect. 50 years
later} (World Scientific, Singapore, 1999).

\bibitem{Bordag1} M. Bordag, U. Mohidden, and V. M. Mostepanenko,
Phys. Rep. {\bf 353}, 1 (2001).

\bibitem{Kirs01} K. Kirsten, {\it Spectral functions in Mathematics
and Physics}. CRC Press, Boca Raton, 2001.

\bibitem{Full73} S. A. Fulling, Phys. Rev {\bf D7}, 2850 (1973).

\bibitem{Full77} S. A. Fulling, J. Phys. A: Math. Gen. {\bf 10},
917 (1977).

\bibitem{Unru76} W. G. Unruh, Phys. Rev. {\bf D14}, 870 (1976).

\bibitem{Boul75} D. G. Boulware, Phys. Rev. {\bf D11}, 1404
(1975).

\bibitem{Avak01} R. M. Avakyan, E. V. Chubaryan, and A. H.
Yeranyan, "'Homogeneous' gravitational field in General
Relativity?", gr-qc/0102030.

\bibitem{Birrell}  N. D. Birrell and P. C. W. Davies, {\it Quantum fields in
curved space} (Chambridge University Press, Cambridge, England,
1982).

\bibitem{Candelas}  P. Candelas and D. Deutsch, Proc. Roy. Soc. Lond. {\bf %
A354}, 79 (1977).

\bibitem{Saharian1}  A. A. Saharian, ''Polarization of the
Fulling--Rindler vacuum by a uniformly accelerated mirror'', hep-th/0110029.

\bibitem{Sahrev}  A. A. Saharian, Izv. Akad. Nauk Arm. SSR. Matematika {\bf 22},
166 (1987)[Sov. J. Contemp. Math. Analysis. {\bf 22}, 70 (1987)];

A. A. Saharian, ''The generalized Abel--Plana formula.
Applications to Bessel functions and Casimir effect'', Report No.
IC/2000/14; hep-th/0002239.

\bibitem{RomSah} A. Romeo and A. A. Saharian, J. Phys. A: Math. Gen. {\bf 35}, 1297 (2002).

\bibitem{Rome01} A. Romeo and A. A. Saharian, Phys. Rev. {\bf
D63}, 105019 (2001).

\bibitem{Saha01} A. A. Saharian, Phys. Rev. {\bf D63}, 125007 (2001).

\bibitem{Reza02} A. H. Rezaeian and A. A. Saharian, Local Casimir
energy for a wedge with a circular outer boundary, hep-th/0110044,
accepted for publication in Class. Quantum Grav.

\bibitem{CandRaine} P. Candelas and D. J. Raine, J. Math. Phys.
{\bf 17}, 2101 (1976).

\bibitem{Davi77} P. C. Davies and S. Fulling, Proc. Roy. Soc.
Lond. {\bf A345}, 59 (1977).

\bibitem{Cand78} P. Candelas and D. Deutsch, Proc. Roy. Soc.
Lond. {\bf A362}, 78 (1978).

\bibitem{Troo79} W. Troost and H. van Dam, Nucl. Phys. {\bf B159},
442 (1979).

\bibitem{Brow85} M. R. Brown and A. C. Ottewill, Phys. Rev. {\bf
D31}, 2514 (1985).

\bibitem{Brow86} M. R. Brown, A. C. Ottewill, and D. Page, Phys. Rev. {\bf
D33}, 2840 (1986).

\bibitem{Hill} C. T. Hill,  Nucl. Phys. B {\bf 277}, 547 (1986).

\bibitem{Frol87} V. P. Frolov and E. M. Serebriany, Phys. Rev. {\bf
D35}, 3779 (1987).

\bibitem{Dowk87} J. S. Dowker, Phys. Rev. {\bf D36}, 3742 (1987).

\bibitem{Pare93} R. Parentani, Class. Quantum Grav. {\bf 10}, 1409
(1993).

\bibitem{More96} V. Moretti and L. Vanzo, Phys. Lett. {\bf B375},
54 (1996).

\bibitem{Taga85} S. Tagaki, Prog. Theor. Phys. {\bf 74}, 142
(1985).

\bibitem{Oogu86} H. Ooguri, Phys. Rev. {\bf D33}, 3573 (1986).

\bibitem{Abramowitz}  M. Abramowitz and I. A. Stegun, {\it Handbook of
Mathematical functions} (National Bureau of Standards, Washington
D.C.,1964).

\bibitem{Amb83}  J. Ambj{\o }rn and S. Wolfram, Ann. Phys. (N.Y.) {\bf
147}, 1 (1983).

\bibitem{Taub51} A. H. Taub, Ann. Math. {\bf 53}, 472 (1951).

\end{thebibliography}
\end{document}